\documentclass[twocolumn]{aastex631}

\newcommand\logAast{\log\tilde{A}_\ast}
\newcommand\logee{\log\epsilon_e}
\newcommand\logeb{\log\epsilon_B}

\received{\today}

\submitjournal{ApJ}

\usepackage[normalem]{ulem}
\usepackage{threeparttable}
\usepackage{amsmath}
\usepackage{pifont}
\usepackage{here}

\newcommand{\cmark}{\ding{51}}%
\newcommand{\xmark}{\ding{55}}%

\shorttitle{MCMC analysis on radio SNe}
\shortauthors{Matsuoka et al.}

\begin{document}

\title{Systematic Investigation into Radio Supernovae with Markov Chain Monte Carlo Analysis:\\ Implications for Massive Stars' Mass Loss and Shock Acceleration Physics}

\author[0000-0002-6916-3559]{Tomoki Matsuoka}
\correspondingauthor{Tomoki Matsuoka}
\email{tmatsuoka@asiaa.sinica.edu.tw}
\affiliation{Institute of Astronomy and Astrophysics, Academia Sinica, No.1, Sec. 4, Roosevelt Road, Taipei 106216, Taiwan}

\author[0000-0003-2611-7269]{Keiichi Maeda}
\affiliation{Department of Astronomy, Kyoto University, Kitashirakawa-oiwakecho, Sakyo-ku, Kyoto 606-8502, Japan}

\author[0000-0003-2579-7266]{Shigeo S. Kimura}
\affiliation{Frontier Research Institute for Interdisciplinary Sciences, Tohoku University, Aramaki-Aoba, Aoba-ku, Sendai 980-8578, Japan}
\affiliation{Astronomical Institute, Graduate School of Science, Tohoku University, Aramaki-Aoba, Aoba-ku, Sendai 980-8578, Japan}

\author[0000-0001-8253-6850]{Masaomi Tanaka}
\affiliation{Astronomical Institute, Graduate School of Science, Tohoku University, Aramaki-Aoba, Aoba-ku, Sendai 980-8578, Japan}

\begin{abstract}
We present a systematic analysis of radio supernovae (SNe) to investigate the statistical tendencies of SN progenitors' mass-loss rates and shock acceleration efficiencies. We conduct parameter estimation through Markov chain Monte Carlo (MCMC) analysis for 32 radio SN samples with a clear peak observed in their light curves, and successfully fit 27 objects with the widely-used radio SN model.
We find the inferred mass-loss rates of stripped-envelope SN progenitors are by an order of magnitude greater ($\sim 10^{-3}\,M_\odot{\rm yr}^{-1}$) than those of SN II progenitors ($<10^{-4}\,M_\odot{\rm yr}^{-1}$).
The efficiencies of electron acceleration and magnetic field amplification are found to be less than $10^{-2}$, and the possibility of their energy equipartition is not ruled out.
On the other hand, we find the following two properties that might be related to limitations of the standard model for radio SNe; one is the extremely high magnetic field amplification efficiency, and the other is the shallower density gradient of the outer ejecta. 
We suggest the new interpretation that these peculiar results are misleading due to the setup that is not included in our model, and we identify the missing setup as a dense CSM in the vicinity of the progenitor. This means that a large fraction of radio SN progenitors might possess dense CSM in the vicinity of the progenitor, which is not smoothly connected with the outer CSM.
\end{abstract}


\section{Introduction}\label{sec:introduction}

Radio emission from core-collapse supernovae (hereafter radio SNe) is one of the electromagnetic signals emitted when the SN shock interacts with circumstellar medium (CSM) \citep[for a review see][]{2017hsn..book..875C}. This process drives acceleration of charged particles such as electrons followed by amplification of turbulent magnetic field.
The accelerated electrons in the magnetic field produce synchrotron emission which we observe in radio band \citep[e.g.,][]{1982ApJ...259..302C}. Thus, modeling radio SNe allows us to constrain the nature of SN-CSM interaction system and shock acceleration physics.

One of the most essential pieces of information provided by radio SNe is the physical structure of the CSM. The peak time and luminosity, as well as the temporal evolution, are largely governed by the magnitude and radial gradient of the CSM density.
Since the information on CSM density profiles can be used to reconstruct the mass-loss history of the SN progenitor prior to core collapse, radio signals from SNe have been adopted as a unique tracer of the final mass-loss activity of massive stars \citep[e.g.,][]{2007ApJ...671.1959W,2017ApJ...835..140M, 2021ApJ...918...34M}.

Another important aspect is plasma microphysics relevant to particle acceleration and magnetic field amplification triggered by an SN shock. Theoretically, a variety of particle-in-cell simulations have raised a possible parameter range of the efficiencies of the electron acceleration and of the amplification of magnetic field \citep[e.g.,][]{2011ApJ...726...75S,PhysRevLett.114.085003, 2015ApJ...798L..28C, 2014ApJ...794...46C, 2015Sci...347..974M, 2017PhRvL.119j5101M}. Construction of an evolutionary model for supernova remnants has also contributed to constraining the nature of particle acceleration \citep{2012ApJ...750..156L}. A series of these studies implies that the fractions of the energy densities of accelerated electrons and amplified magnetic field range from $10^{-5}$ to $10^{-1}$ in thermal energy density. Gamma-ray burst afterglow also provides us with the opportunities to give constraints on the efficiency parameters, though the shock wave in gamma-ray burst jets are considered to be in a relativistic regime \citep[e.g.,][]{2005ApJ...627..861E,2014ApJ...785...29S,2014MNRAS.442.3147B, 2017MNRAS.472.3161B, 2023MNRAS.518.1522D, 2021MNRAS.504.5647S}.

Many attempts at fitting individual radio SN light curves have even been carried out previously (see Table~\ref{tab:sample} for reference).
However, a systematic investigation into comprehensive samples of radio SNe within a single model framework has been largely missing, unlike optical light-curve analyses \citep[e.g.,][]{2016MNRAS.457..328L,2022A&A...660A..40M,2022A&A...660A..41M,2022A&A...660A..42M,2024ApJ...969...57S}.
Using the results of previous papers focusing on individual radio SN objects to address possible systematic tendencies has a critical drawback, since there have been differences in the fitting strategies and in the selection of parameters to be surveyed or fixed \citep[for example see Appendix in][]{2022ApJ...938...84D}.
Observationally, \citet{2021ApJ...908...75B} has recently compiled almost all radio observations of core-collapse SNe conducted ever. They investigated the relationship between peak luminosity and peak time depending on SN types, whereas only the mean values of the mass-loss rate were estimated for samples of SN~II and SN~Ibc based on simple fitting formulae with several assumptions.
Under these circumstances, an important step is to conduct a systematic survey of radio SN properties based both on a systematic sample and a unified/single model framework. 

In this paper, we conduct a systematic investigation of radio SNe by using a Markov chain Monte Carlo (MCMC) analysis. We employ a widely-employed model to calculate radio luminosity of SNe, which allows us to directly compare the fitting values of parameters and to infer systematic tendencies. This paper is organized as follows. In Section \ref{sec:sample}, we show the selection procedure of our radio SN samples treated in this study. In Section \ref{sec:method}, we describe our model for radio SNe and the procedure of the MCMC analysis. In Section \ref{sec:fitting}, we present the overall fitting results reproduced by our parameter estimation. Further analysis, discussion, and caveats regarding CSM, shock acceleration physics, and ejecta structure are described in Section \ref{sec:Discussion_CSM}, \ref{sec:Discussion_PAMFA}, and \ref{sec:Discussion_ejecta}, respectively. Finally, the contents of this paper are summarized in Section \ref{sec:summary}.

\section{Sample selection}\label{sec:sample}
\cite{2021ApJ...908...75B} presented a comprehensive data set of radio SNe observed ever.
It is reported therein that observations of 294 radio SNe have been ever conducted mainly in the wavelength of the centimeter, and the significant detections have been confirmed for 90 SNe, corresponding to $\sim30$\% of all SNe for which radio observations have been conducted. We look through the published data of 90 radio SNe, and construct the radio SN samples based on the following conditions; (i) the radio data set has been published, and (ii) there is a series of radio detection data showing a clear peak in the light curve. As a result, we select 30 radio SNe from the catalog in \cite{2021ApJ...908...75B}. In addition, we include the radio data set of SN~2016X \citep{2022A&A...666A..82R} and SN~2016gkg \citep{2022ApJ...934..186N} whose observations have been recently reported. In total, these 32 radio SNe with a high-quality radio data set form the golden sample for our analysis. We classify the SN samples into SNe II, IIb, Ib, Ic, and broad-lined Ic (IcBL).
We note that SNe~IIn are not treated in our study because the long-term evolution of SNe IIn makes it difficult to obtain sufficient data set involving a clear peak \citep[see e.g.,][]{2015ApJ...810...32C}. Although we include SN~2014C in our samples, which has exhibited metamorphosis from SN Ib to SN IIn, we only treat the data obtained within 200~days after the explosion (see also Section~\ref{sec:data_removal}).

Table~\ref{tab:sample} summarizes the SN types, ejecta masses ($M_{\rm ej}$), kinetic energies ($E_{\rm ej}$), and the distances ($D$) of our 32 radio SN samples, all of which are necessary to calculate radio luminosity from SNe. $M_{\rm ej}$ and $E_{\rm ej}$ affect the velocity of the SN shock and were determined by the optical bolometric light curve fitting in most previous papers.
$D$ determines the normalization of the observed flux density of the radio emission. In this study, we just fix them to the values declared in the previous studies. Some stripped-envelope SN (SESN) samples have no estimate of $M_{\rm ej}$ and $E_{\rm ej}$ in previous studies, for the case of which we adopt the representative parameters proposed in \cite{2016MNRAS.457..328L}.

\begin{table*}
    \centering
	\caption{Our radio SN samples and their properties.}\label{tab:sample}
	\begin{tabular}{lcccccccc} 
		\hline
		\hline
		SN type & SN name & $M_{\rm ej}\,(M_\odot)$ & $E_{\rm ej}\,(10^{51}\,{\rm erg})$ & $D\,({\rm Mpc})$ & References of & References of & References of & $\chi_{\rm red}^2<5$ \\
		 & & & & & $M_{\rm ej}, E_{\rm ej}$ \tnote{1} & $L_{\rm bol}$ \tnote{1} & radio data\tnote{1} & \\
		\hline
II & SN 1987A & 14.0 & 1.1 & 0.051 & 1 & 2 & 3 & \cmark \\
& SN 2004dj & 10.0 & 0.75 & 3.4 & 4 & 4 & 5 & \cmark \\
& SN 2012aw & 20.0 & 1.5 & 10.0 & 6 & 6 & 7 & \cmark \\
& SN 2016X & 28.0 & 1.7 & 15.2 & 8 & 9 & 10 & \cmark \\
\hline
IIb & SN 1993J & 2.7 & 1.0 & 3.7 & 11 & 11 & 12 & \cmark \\
& SN 2001gd & 2.2$^\dagger$ & 1.0$^\dagger$ & 17.5 & 13 & 13 & 14 & \cmark \\
& SN 2001ig & 2.2$^\dagger$ & 1.0$^\dagger$ & 9.3  & 13 & 13 & 15 & \cmark \\
& SN 2011dh & 3.0 & 1.0 & 7.9 & 16 & 16 & 17, 18 & \cmark \\
& SN 2011ei & 1.6 & 2.5 & 28.7 & 19 & 19 & 19 & \cmark \\
& SN 2011hs & 1.8 & 0.8 & 21.3 & 20 & 20 & 20 & \xmark \\
& SN 2013df & 1.1 & 0.8 & 18.1  & 21 & 21 & 22 & \xmark \\
& SN 2016gkg & 3.4 & 1.2 & 26.4 & 23 & 24 & 25 & \cmark \\
\hline
Ib & SN 1983N & 3.0 & 1.0 & 4.9 & 26 & 27 & 27 & \cmark \\
& SN 2004dk & 3.7 & 1.8 & 20.8 & 13\tnote{4} & 13\tnote{4} & 28 & \cmark \\
& SN 2004gq & 1.8 & 1.9 & 24.3 & 13\tnote{4} & 13\tnote{4} & 28 & \cmark \\
& SN 2007gr & 2.7 & 2.5 & 5.2 & 29 & 29 & 30 & \cmark \\
& SN 2007uy & 4.4 & 15.0 & 28.6 & 31 & 31 & 32 & \cmark \\
& SN 2008D & 4.0 & 3.0 & 28.6 & 33 & 33 & 34 & \xmark \\
& SN 2012au & 4.0 & 10.0 & 22.9 & 35 & 35 & 36 & \cmark \\
& AT2014ge & 2.6$^\dagger$ & 1.6$^\dagger$ & 15.5 & 13 & 13 & 37 & \cmark \\
& SN 2014C & 1.7 & 1.7 & 14.7 & 38 & 38 & 38 & \cmark \\
\hline
Ic & SN 1990B & 3.0$^\dagger$ & 1.9$^\dagger$ & 17.4 & 13 & 13 & 39 & \cmark \\
& SN 1994I & 1.0 & 1 & 8.4 & 40 & 41 & 42 & \cmark \\
& SN 2003L & 3.0$^\dagger$ & 1.9$^\dagger$ & 92.0 & 13 & 13 & 43 & \cmark \\
& SN 2004cc & 3.0$^\dagger$ & 1.9$^\dagger$ & 17.4 & 13 & 13 & 28 & \xmark \\
& SN 2020oi & 1.0 & 1.0 & 15.0 & 44 & 45 & 45, 46 & \cmark \\
\hline
IcBL & SN 2002ap & 3.8 & 7.0 & 8.9 & 47\tnote{2} & 47 & 48 & \cmark \\
& SN 2003bg & 4.8 & 5.0 & 19.3 & 49 & 49 & 50 & \cmark \\
& SN 2007bg & 1.5 & 4.0 & 155.0 & 51 & 51 & 52 & \cmark \\
& SN 2009bb  & 4.1 & 18.0 & 43.5 & 53 & 53 & 54 & \xmark \\
& SN 2012ap & 2.7 & 9.0 & 40.0 & 55 & 55 & 56 & \cmark \\
& SN 2016coi & 5.5 & 7.0 & 18.1 & 57 & 58 & 57 & \cmark \\
		\hline
		\hline
\end{tabular}
\begin{tablenotes}
\item[1] References. (1) \cite{2000ApJ...532.1132B}, (2) \cite{1990AJ.....99..650S}, (3) \cite{1987Natur.327...38T}, (4) \cite{2006AJ....131.2245Z}, (5) \cite{2018ApJ...863..163N}, (6) \cite{2014ApJ...787..139D}, (7) \cite{2014ApJ...782...30Y}, (8) \cite{2019MNRAS.490.2042U}, (9) \cite{2018MNRAS.475.3959H}, (10) \cite{2022A&A...666A..82R}, (11) \cite{1994ApJ...420..341S}, (12) \cite{2007ApJ...671.1959W}, (13) \cite{2016MNRAS.457..328L}, (14) \cite{2007ApJ...671..689S}, (15) \cite{2004MNRAS.349.1093R}, (16) \cite{2012ApJ...757...31B}, (17) \cite{2012ApJ...750L..40K}, (18) \cite{2012ApJ...752...78S}, (19) \cite{2013ApJ...767...71M}, (20) \cite{2014MNRAS.439.1807B}, (21) \cite{2014MNRAS.445.1647M}, (22) \cite{2016ApJ...818..111K}, (23) \cite{2018Natur.554..497B}, (24) \cite{2019MNRAS.485.1559P}, (25) \cite{2022ApJ...934..186N}, (26) \cite{1986ApJ...301..790W}, (27) \cite{1990ApJ...361L..23S}, (28) \cite{2012ApJ...752...17W}, (29) \cite{2009A&A...508..371H}, (30) \cite{2010ApJ...725..922S}, (31) \cite{2013MNRAS.434.2032R}, (32) \cite{2011ApJ...726...99V}, (33) \cite{2009ApJ...692.1131T}, (34) \cite{2008Natur.453..469S}, (35) \cite{2013ApJ...770L..38M}, (36) \cite{2014ApJ...797....2K}, (37) \cite{2019ApJ...877...79C}, (38) \cite{2017ApJ...835..140M}, (39) \cite{1993ApJ...409..162V}, (40) \cite{1994Natur.371..227N}, (41) \cite{1996AJ....111..327R}, (42) \cite{2011ApJ...740...79W}, (43) \cite{2005ApJ...621..908S}, (44) \cite{2021ApJ...908..232R}, (45) \cite{2020ApJ...903..132H}, (46) \cite{2021ApJ...918...34M}, (47) \cite{2002ApJ...572L..61M}, (48) \cite{2002ApJ...577L...5B}, (49) \cite{2009ApJ...703.1624M}, (50) \cite{2006ApJ...651.1005S}, (51) \cite{2010A&A...512A..70Y}, (52) \cite{2013MNRAS.428.1207S}, (53) \cite{2011ApJ...728...14P}, (54) \cite{2010Natur.463..513S}, (55) \cite{2015ApJ...799...51M}, (56) \cite{2015ApJ...805..187C}, (57) \cite{2019ApJ...883..147T}, (58) \cite{2018MNRAS.478.4162P}
\item[2] a. We adopt the average value of $M_{\rm ej}$ proposed in \citet{2002ApJ...572L..61M}.
\item[3] b. The values with $\dagger$ represent the mean value derived in \citet{2016MNRAS.457..328L}.
\item[4] c. $M_{\rm ej}$ and $E_{\rm ej}$ of SN\,2004dk and SN\,2004gq refers to the values proposed in Table\,5 of \citet{2016MNRAS.457..328L}.
\item[5] d. The final column shows whether the reduced chi-square is below 5, which means the successful fitting. See Section~\ref{sec:MCMC}.
\end{tablenotes}
\end{table*}

\section{Model for radio supernovae and MCMC analysis}\label{sec:method}

\subsection{Radio emission from SNe}\label{sec:radioSN}

In this subsection, we describe the analytical model for radio SNe. Since we employ the widely-used procedure for computing radio luminosity, here we briefly describe the content of the model and explicitly give expressions that involve parameters we will survey in this study. For the details in our radio SN model we refer readers to the relevant studies \citep[e.g,.][]{1982ApJ...259..302C,1998ApJ...499..810C,1998ApJ...509..861F,2006ApJ...641.1029C,2006ApJ...651..381C,2012ApJ...758...81M,2017hsn..book..875C,2019ApJ...885...41M,2020ApJ...898..158M,2021ApJ...918...34M,2022ApJ...930..143M,2024ApJ...960...70M}

We consider an SN ejecta experiencing homologous expansion and the density profile of the SN ejecta consists of a flat inner component and a steep outer component. We consider the model in the expansion of the SN shock to be driven by the ram pressure from the outer ejecta. The density structure is given as follows:
\label{eq:vejrhoej}\begin{eqnarray}
\rho_{\rm ej} &=& 
\frac{3M_{\rm ej}}{4\pi v_c^3}
\frac{n-3}{n} t^{-3}
\left(\frac{v}{v_c}\right)^{-n},\\
v_c &=& \left(\frac{10E_{\rm ej}}{3M_{\rm ej}} \frac{n-5}{n-3}\right)^{1/2},
\end{eqnarray}
where $n$ specifies the density gradient of the outer SN ejecta and is limited in the range of $n>5$ \citep{1982ApJ...258..790C}. 
We also consider the power-law distribution of CSM density profile described as follows:
\begin{eqnarray}\label{eq:rhocsm}
\rho_{\rm CSM} = \mathcal{D} r^{-s} = 5\times 10^{19} \tilde{A}_\ast \left(\frac{r}{10^{15}\,{\rm cm}}\right)^{-s},
\end{eqnarray}
where $\tilde{A}_\ast$ quantifies the CSM density scale at the radius of $r=10^{15}\,{\rm cm}$. $0<s<3$ is required in order to realize the decelerating SN shock. The value $s=2$ describes the steady wind, and
\begin{eqnarray}\label{eq:qA}
\tilde{A}_\ast = \frac{\dot{M}}{4\pi v_w}\times \frac{1}{5\times 10^{11}\,{\rm g\ cm}^{-1}}
\end{eqnarray}
is deduced. We mention that if we focus on $s=2$ then $\tilde{A}_\ast$ is equivalent with $A_\ast$ used in the literature \cite[e.g.,][]{2012ApJ...758...81M, 2018MNRAS.478..110S}; for typical Wolf-Rayet stars with steady wind ($v_w = 1000\,{\rm km}\,{\rm s}^{-1}$), $\tilde{A}_\ast = A_\ast\sim1$ is expected \citep{2006ApJ...651..381C}.

We adopt the thin-shell approximation to describe the time evolution of an SN shock \citep[][but see also Section\,\ref{sec:Discussion_ejecta} for the validity of this approximation]{1982ApJ...259..302C}. Specifically, we use the equations of the radius $R_{\rm sh}$ and the velocity $V_{\rm sh}$ of the SN shock as follows:
\begin{eqnarray}\label{eq:Rsh}
R_{\rm sh} &=& \left[\frac{(3-s)(4-s)}{(n-3)(n-4)}\frac{U_c^n}{\mathcal{D}}\right]^{1/(n-s)}
t^{(n-3)/(n-s)}, \\
V_{\rm sh} &=& \frac{dR_{\rm sh}}{dt} = \frac{n-3}{n-s}\frac{R_{\rm sh}}{t},\label{eq:Vsh}
\end{eqnarray}
where $U_c$ is given as follows:
\begin{eqnarray}\label{eq:Ucn}
U_c &=& \left(\frac{3M_{\rm ej}}{4\pi v_c^{3-n}}\frac{n-3}{n}\right)^{1/n}.
\end{eqnarray}

Next, we move to the parametrization of particle acceleration and magnetic field amplification.
We compute the energy densities of electrons and magnetic field assuming that these are carried from the kinetic energy of the SN shock as follows:
\begin{eqnarray}
u_e &=& f_{\rm sh} \epsilon_e \rho_{\rm CSM}(r=R_{\rm sh}) V_{\rm sh}^2, \label{eq:ue} \\
u_B &=& \frac{B^2}{8\pi} = f_{\rm sh} \epsilon_B \rho_{\rm CSM}(r=R_{\rm sh}) V_{\rm sh}^2, \label{eq:uB}
\end{eqnarray}
where we use $f_{\rm sh}=9/8$ taking into account the shock compression factor and the relative velocity between upstream and downstream \citep[][and see also Section \ref{sec:Discussion_ejecta}]{2016MNRAS.460...44P}.
From Equation (\ref{eq:uB}) we can estimate the strength of the amplified magnetic field. Equation (\ref{eq:ue}) will be used to quantify the normalization of the SED of relativistic electrons $N(\gamma)$. Assuming that there is a balance between injection and cooling of relativistic electrons at any time, we can derive analytical formalization of $N(\gamma)$ as follows \citep{1998ApJ...509..861F}:
\begin{eqnarray}
N(\gamma) = \frac{p-2}{p-1}\frac{u_e V_{\rm sh} t_{\rm cool}(\gamma)}{\gamma_m^{2-p}m_e c^2 \Delta R_{\rm sh}} \gamma^{-p}, \label{eq:Ngamma}
\end{eqnarray}
where $p$ is the spectral index of the relativistic electrons.
$t_{\rm cool}(\gamma)$ is the cooling timescale as a function of $\gamma$ in which four kinds of cooling processes are taken into account; synchrotron cooling, inverse Compton cooling, adiabatic expansion, and Coulomb interaction, though the last one can hardly affect the results \citep{1998ApJ...509..861F, 2019ApJ...885...41M, 2021ApJ...918...34M, 2022ApJ...936...98B}.
We set the minimum Lorentz factor of electrons to $\gamma_m=2$ because the velocity of the SN shock is in a non-relativistic regime (\citealp{2023MNRAS.524.6004W}, but see also \citealp[][]{2024ApJ...960...70M}). $\Delta R_{\rm sh}$ is the geometrical thickness of the shocked CSM region, which is a few tens of percent of $R_{\rm sh}$ depending on the choice of $n$ \citep{1982ApJ...258..790C,1994ApJ...420..268C,1999ApJS..120..299T}.

By using the computed electron spectrum $N(\gamma)$ and the magnetic field $B$ we can estimate the synchrotron emissivity $j_{\nu, {\rm syn}}$, absorption coefficient of synchrotron self-absorption $\alpha_{\nu, {\rm syn}}$, and the corresponding optical depth $\tau_{\nu, {\rm syn}}$. We follow the formalisms described in the literature to calculate these quantities relevant to synchrotron emission \citep[e.g.,][]{2020ApJ...898..158M,2024ApJ...960...70M}. The dependence of synchrotron emission power on the pitch angle of emitter electrons is absorbed in the average of the solid angle via the formula in \cite{2010PhRvD..82d3002A}.
 
At last, we can estimate the radio SN luminosity as functions of time ($t$) and frequency ($\nu$) as follows:
\begin{eqnarray}\label{eq:flux}
F_{\rm model}(t, \nu; \vec{\theta}) = \pi \frac{R_{\rm sh}^2}{D^2}\frac{j_{\nu, {\rm syn}}}{\alpha_{\nu, {\rm syn}}}(1-e^{-\tau_{\nu, {\rm syn}}}),
\end{eqnarray}
where $\vec{\theta}=(\tilde{A}_\ast, \epsilon_e, \epsilon_B, p, s, n)$ denotes the vector of parameters we will survey in our MCMC sampling procedure.

\subsection{Bayesian framework and parameter estimation}\label{sec:MCMC}
We employ the python package \verb'emcee' \citep{2013PASP..125..306F} to find out the parameter set through the fit to the observed data of each radio SN. \verb'emcee' is capable of performing MCMC sampler with many walkers for a large number of iterations and of suggesting the probability density function in the surveyed parameters. For one object of a radio SN, we prepare 100 walkers and iterate the model calculations for $10^5$ times for each walker, and finally discard the first 5000 steps of calculation result to exclude the samples before the burn-in phase. Furthermore, we extract the sample set from the iterated data with the distance of $100$ steps in order to realize the random sampling.

The plausible sets of the parameters are searched in order for the posterior distribution $P(\vec{\theta} \mid \vec{F}_{\rm obs})$ (in other words, the conditional probability that the parameter set $\vec{\theta}$ should be realized given the observational data $\vec{F}_{\rm obs}$) to be maximized, which is defined as follows:
\begin{eqnarray}\label{eq:posterior}
P(\vec{\theta} \mid \vec{F}_{\rm obs}) = \frac{P( \vec{F}_{\rm obs} \mid \vec{\theta}) \cdot P(\vec{\theta})}{Z(\vec{F}_{\rm obs})},
\end{eqnarray}
where $P(\vec{F}_{\rm obs} \mid \vec{\theta})$ is the likelihood function, $P(\vec{\theta})$ is the prior distribution of the parameters, and $Z(\vec{F}_{\rm obs})$ denotes the normalization factor dependent on only the data set of the radio SN. The likelihood function is defined by the product of the residual between the observed and computed radio flux density for the number of observational data $N_{\rm obs}$ as follows \citep[e.g.,][]{2024JHEAp..41....1O}:
\begin{eqnarray}\label{eq:likelihood}
&&P(\vec{F}_{\rm obs} \mid \vec{\theta}) \nonumber \\
&&= \prod_{i=1}^{N_{\rm obs}} \frac{1}{\sqrt{2\pi \sigma_{{\rm obs},i}^2}} \exp\left(-\frac{(F_{{\rm obs},i} - F_{{\rm model}}(t_i,\nu_i;\vec{\theta}))^2}{2\sigma_{{\rm obs},i}^2} \right),\nonumber \\
\end{eqnarray}
where $\sigma_{{\rm obs},i}$ is the observed error associated with the $i$th data of the detected radio flux density ($F_{{\rm obs},i}$). The observed errors mainly consist of intrinsic and systematic errors, and the latter may depend on the method to analyze the raw data as well as the observational conditions, the way of calibrations, etc, implying that even $\sigma_{{\rm obs},i}$ may involve relevant uncertainties.  For instance, Table 1 in \cite{2014ApJ...797....2K} has listed only the flux density error associated with the image rms for SN 2012au, though the systematic error is likely to be taken into account in their main figures and discussion. In this study, we basically adopt the values of the error published in each paper, and we additionally impose a lower limit of $\sigma_{{\rm obs},i} = 0.1 F_{{\rm obs},i}$, which is a typical error in flux calibration.

We consider uniform distributions as prior functions of parameters in $\vec{\theta}$ with the lower and upper boundaries given as shown in Table \ref{tab:params_priors}. We note that $\tilde{A}_\ast, \epsilon_e$, and $\epsilon_B$ are treated on a logarithmic scale to accommodate variations across several orders of magnitude. In addition, we impose the constraint of the number fraction of relativistic electrons to the prior function. As discussed in \citet{2024ApJ...960...70M}, the minimum Lorentz factor of electrons $\gamma_m$ is indeed described as a function of the number fraction of relativistic electrons with respective to thermal electrons ($f_e$), as well as $\epsilon_e$ and $V_{\rm sh}$. As we fix the value of $\gamma_m$ as 2, the combination of large $\epsilon_e$ and fast shock velocity may lead to the unphysical situation of $f_e > 1$. To avoid this, we additionally set the prior function to zero once the parameter set gives $f_e > 1$.

\begin{table*}
	\centering
	\begin{threeparttable}
	\caption{The list of surveyed parameters in this study and their prior functions}\label{tab:params_priors}
	\begin{tabular}{ccc} 
		\hline
		\hline
		 Parameter & Specification  & Prior bound \\
		\hline
		$\log\tilde{A}_\ast$ & CSM density scale & $(-2,4)$ \\
		$\log\epsilon_e$ & Electron acceleration efficiency & $(-5,0)$ \\
		$\log\epsilon_B$ & Magnetic field amplification efficiency & $(-5,0)$ \\
		$p$ & Spectral index of relativistic electrons & $(2,4)$ \\
		$s$ & Density slope of the CSM & $(0,3)$ \\
		$n$ & Density slope of the outer ejecta & $(5,30)$ \\
		\hline
		\hline
	\end{tabular}
	\end{threeparttable}
\end{table*}

Finally, to quantitatively evaluate the degree of fitting, we estimate the reduced chi-square $\chi_{\rm red}^2$ for the most plausible parameter set defined as follows:
\begin{eqnarray}
    \chi_{\rm red}^2 = \frac{1}{N_{\rm obs}-\Vert\vec{\theta}\Vert}\sum_{i=1}^{N_{\rm obs}} \frac{(F_{{\rm obs},i} - F_{\rm model}(t_i,\nu_i;\vec{\theta}))^2}{\sigma_{{\rm obs},i}^2}.\nonumber \\
\end{eqnarray}
The radio SN samples with $\chi_{\rm red}^2 < 5$ are proceeded to the subsequent analysis in Sections\,\ref{sec:fitting}, \ref{sec:Discussion_CSM}, and \ref{sec:Discussion_PAMFA}.

\subsection{Treatment of observational data}\label{sec:data_removal}
Before we showcase our MCMC parameter estimation results, we note that in our simulations we exclude the observational data in the first {30}\,days since the explosion.
Recently it has been indicated that the circumstellar environment in the vicinity of SN progenitors is deviated from the extrapolation of the CSM residing in the outer region. For example, some SNe II exhibit flash spectroscopic features characterized by highly-ionized narrow emission lines laid on the blue continuum \citep[e.g.,][]{2017NatPh..13..510Y, 2022ApJ...924...15J, 2023ApJ...954L..42J,
2024Natur.627..759Z, 2024MNRAS.528.4209M}, implying the dense CSM component residing in the vicinity of the SN progenitor. It is also reported that even an SN Ic\,2020oi possesses an inhomogeneous CSM density structure at the length scale of $r\sim10^{15}\,{\rm cm}$ \citep{2021ApJ...918...34M}. Our radio SN model is not capable of dealing with the inhomogeneous density distribution because of the assumption of a single-component CSM (see Equation \ref{eq:rhocsm}). Therefore we will not attempt to fit the observational data in the early phase. This means that the parameters we survey in this paper are intrinsic to the nature of the widespread CSM configuration outside $r\gtrsim 10^{15}\,{\rm cm}$ (see Section\,\ref{subsec:CSM_largescale}). It is worth mentioning that this early-phase data exclusion could also remove the potential effect of free-free absorption which is not included in our model but can be important if the CSM density is high \citep{1988AA...192..221L,2006ApJ...641.1029C,2019ApJ...885...41M}. We also note that if there is an inhomogeneous CSM density structure developed in the vicinity of the SN progenitor, the time evolution of the SN shock is firstly influenced by the inner CSM, and the subsequent hydrodynamical properties including the SN shock would also be affected \citep{2025arXiv250414255M}. This effect is not taken into consideration in our current model.

Due to the limitation of our analytical model for radio SNe, we focus on the time range during which the radio light curves are characterized by a single peak with gradual rise and decline phases. We excluded observational data showing non-standard characteristics, specifically: double-peaked light curves and unusually rapid rise or decay of radio luminosity. We refer readers to Appendix\,\ref{app:anomaly} for the detailed descriptions of the peculiar behavior of light curves seen in some radio SN samples.
Furthermore, Section~\ref{subsec:CSM_vicinity} discusses how our parameter estimation results are affected if the early-phase data, as omitted in our main analysis, are included (see also Appendix~\ref{app:allepoch}).

\section{Fitting results}\label{sec:fitting}

\begin{figure*}
\centering
\includegraphics[width=0.7\linewidth]{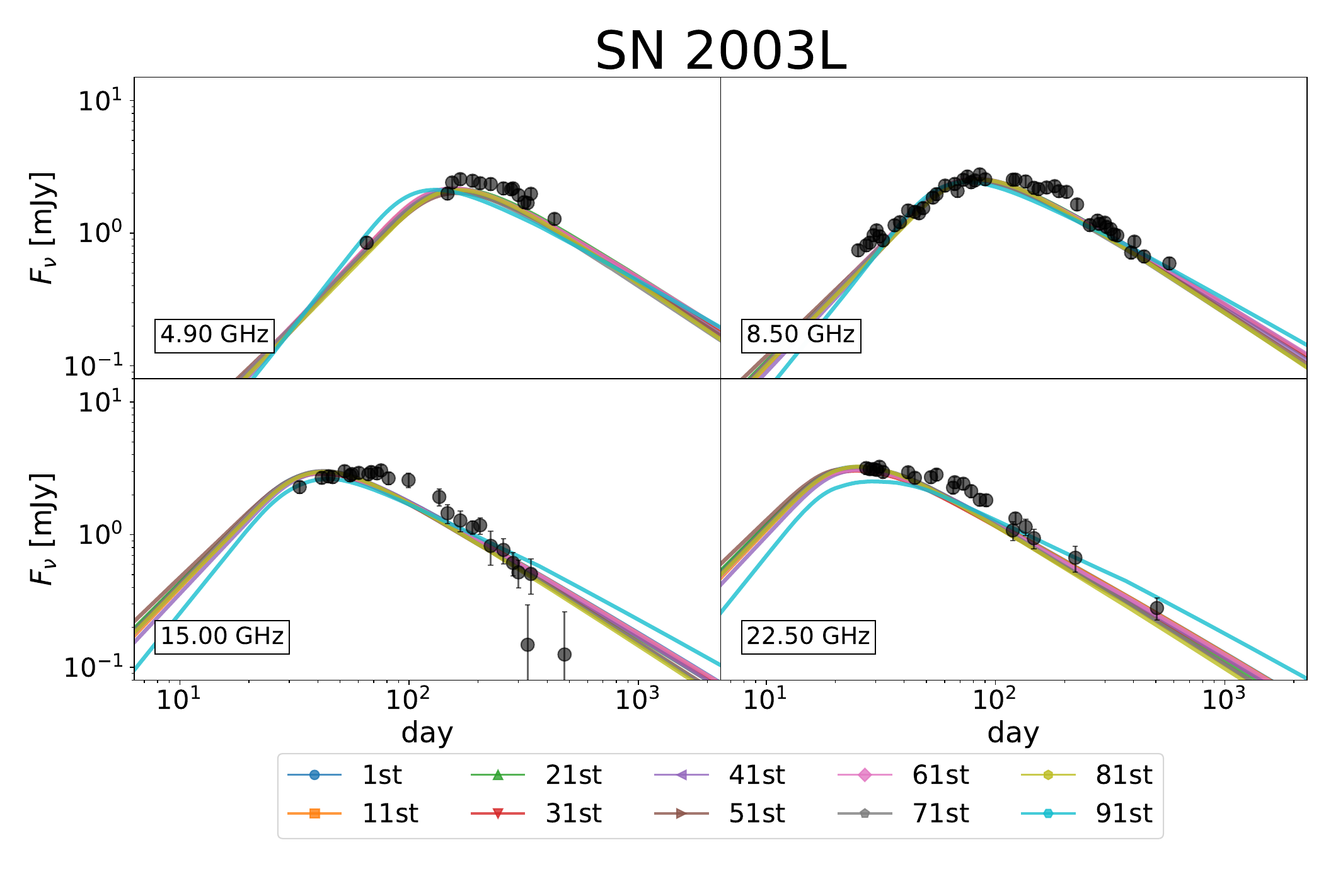}
\includegraphics[width=0.7\linewidth]{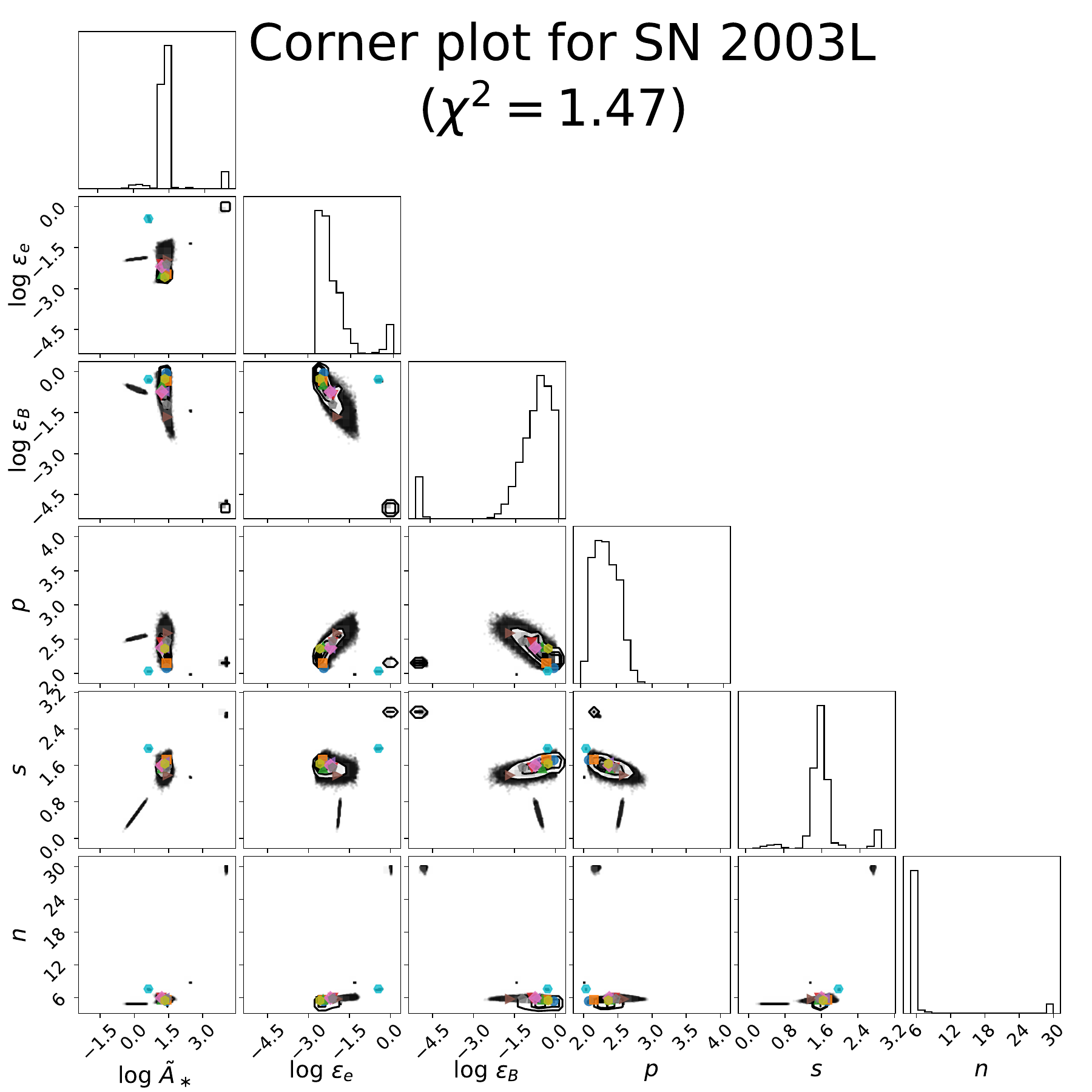}
\label{fig:sn2003L}
\caption{Top: Multi-band radio light curves of SN~2003L. Bottom: A corner plot of the marginalized posterior probability distribution of the parameters in SN~2003L. Different colors of light curves and points correspond to different ranks of chi-square values denoted in the legend.}
\end{figure*}

After the inspection of $\chi^2_{\rm red}$, we find that 27 radio SN samples satisfy the condition of $\chi_{\rm red}^2<5$, representing the well-fitted model. These samples are noted by the checkmark in Table~\ref{tab:sample} and filed into the subsequent analysis such as constructing the probability density functions for parameters in $\vec{\theta}$. Figure~1 shows an example of a well-fitted radio light curve SN~2003L and the marginalized posterior probability distribution of the parameters suggested by our MCMC parameter estimation. Tables\,\ref{tab:median_1sigma_eachSN_logAlogeelogeB} and \ref{tab:median_1sigma_eachSN_psn} show the modes and medians of surveyed parameters with the associated $1\sigma$ credible intervals which are derived as the range of each parameter that covers the 16th to 84th percentiles in the 1D marginalized distribution.

\begin{figure*}[htbp]
\includegraphics[width=0.5\linewidth]{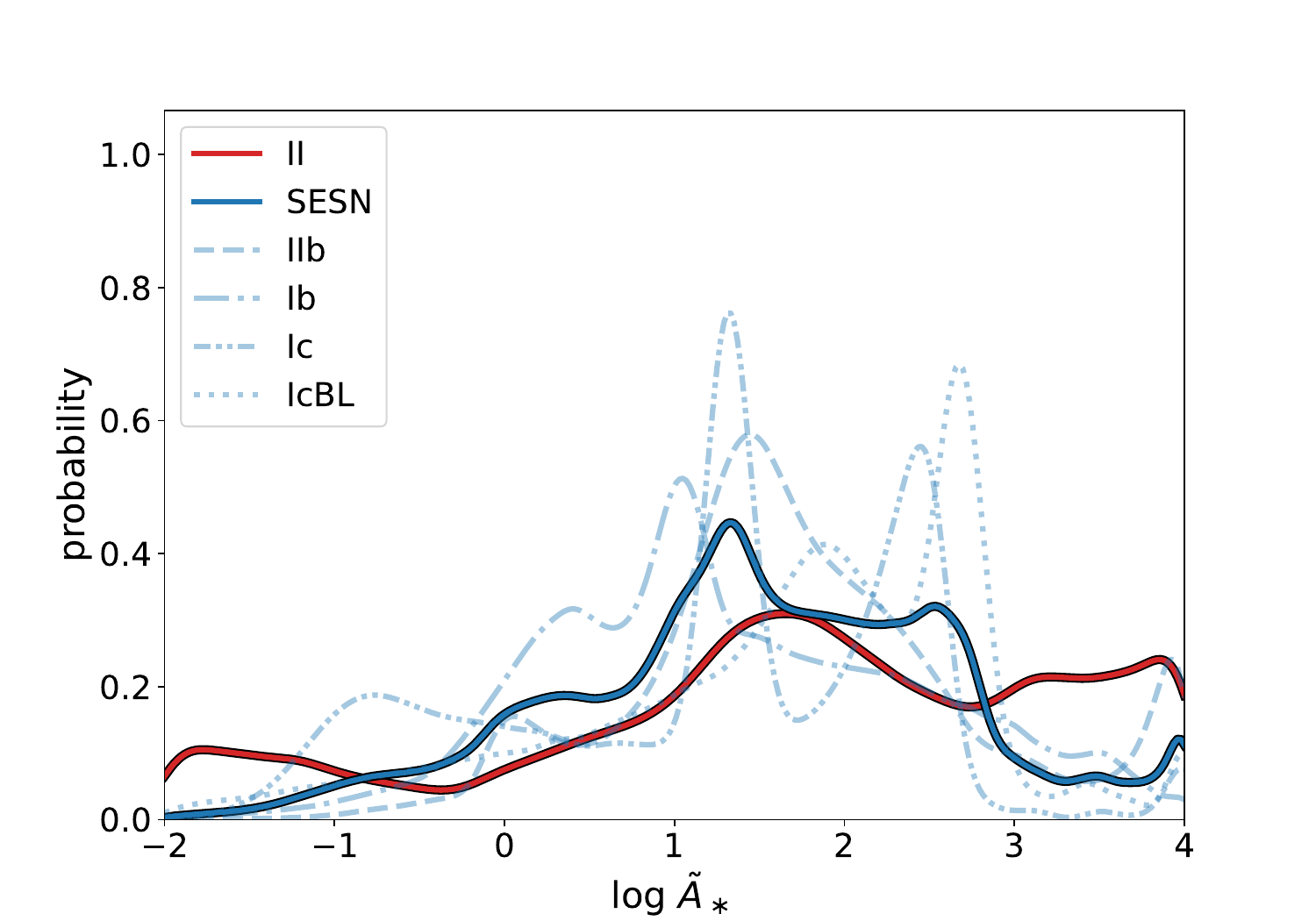}
\includegraphics[width=0.5\linewidth]{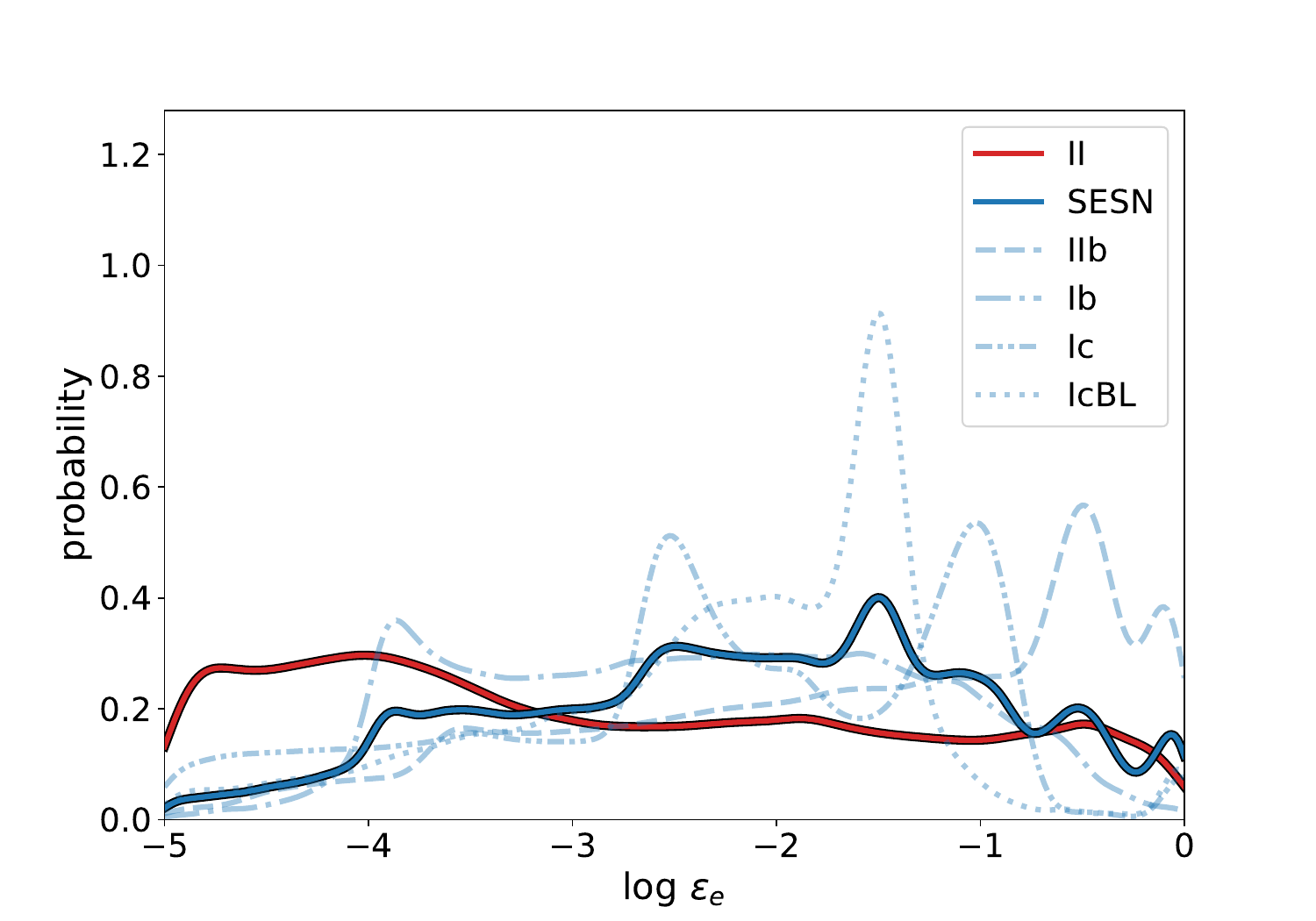} \\
\includegraphics[width=0.5\linewidth]{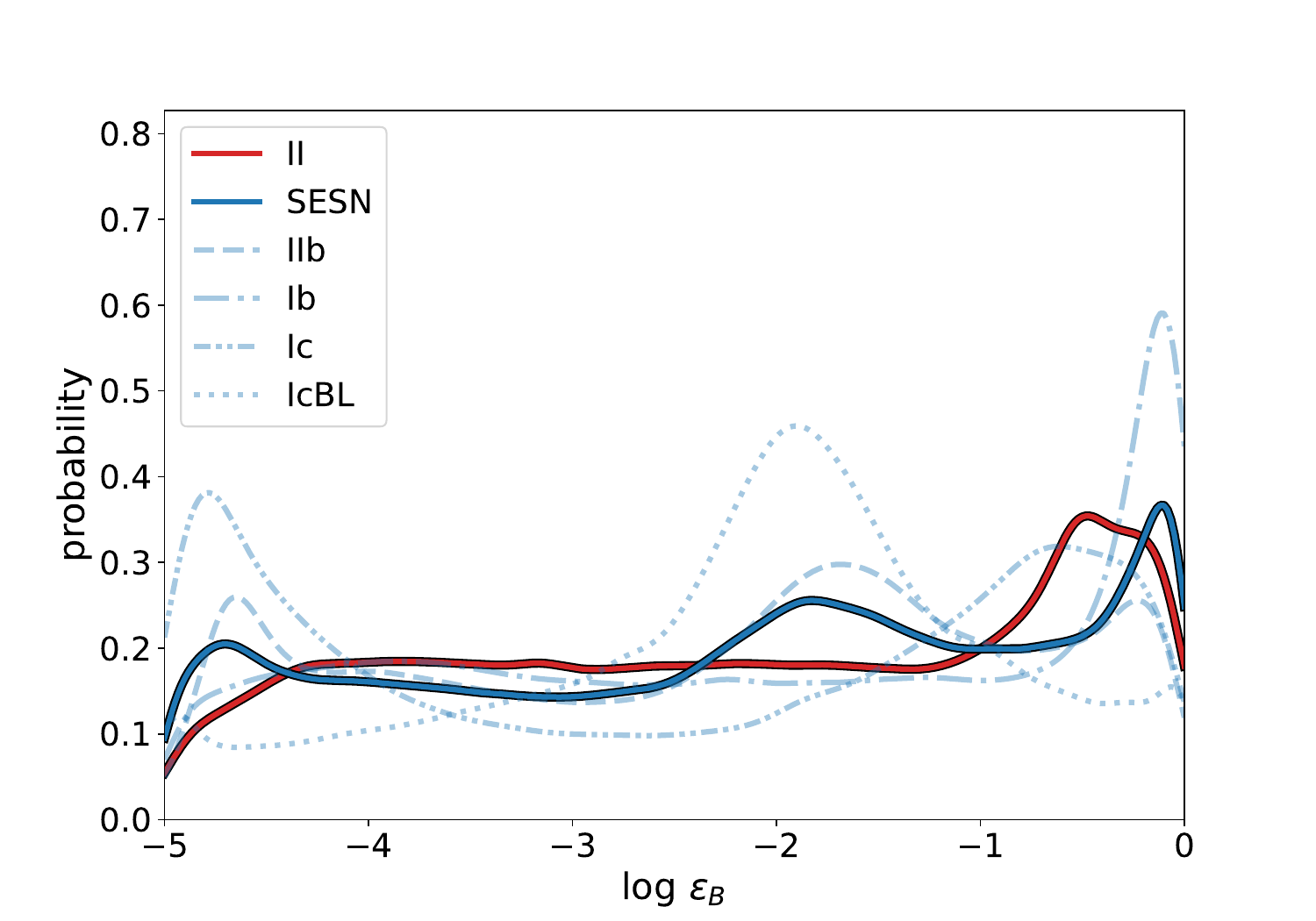}
\includegraphics[width=0.5\linewidth]{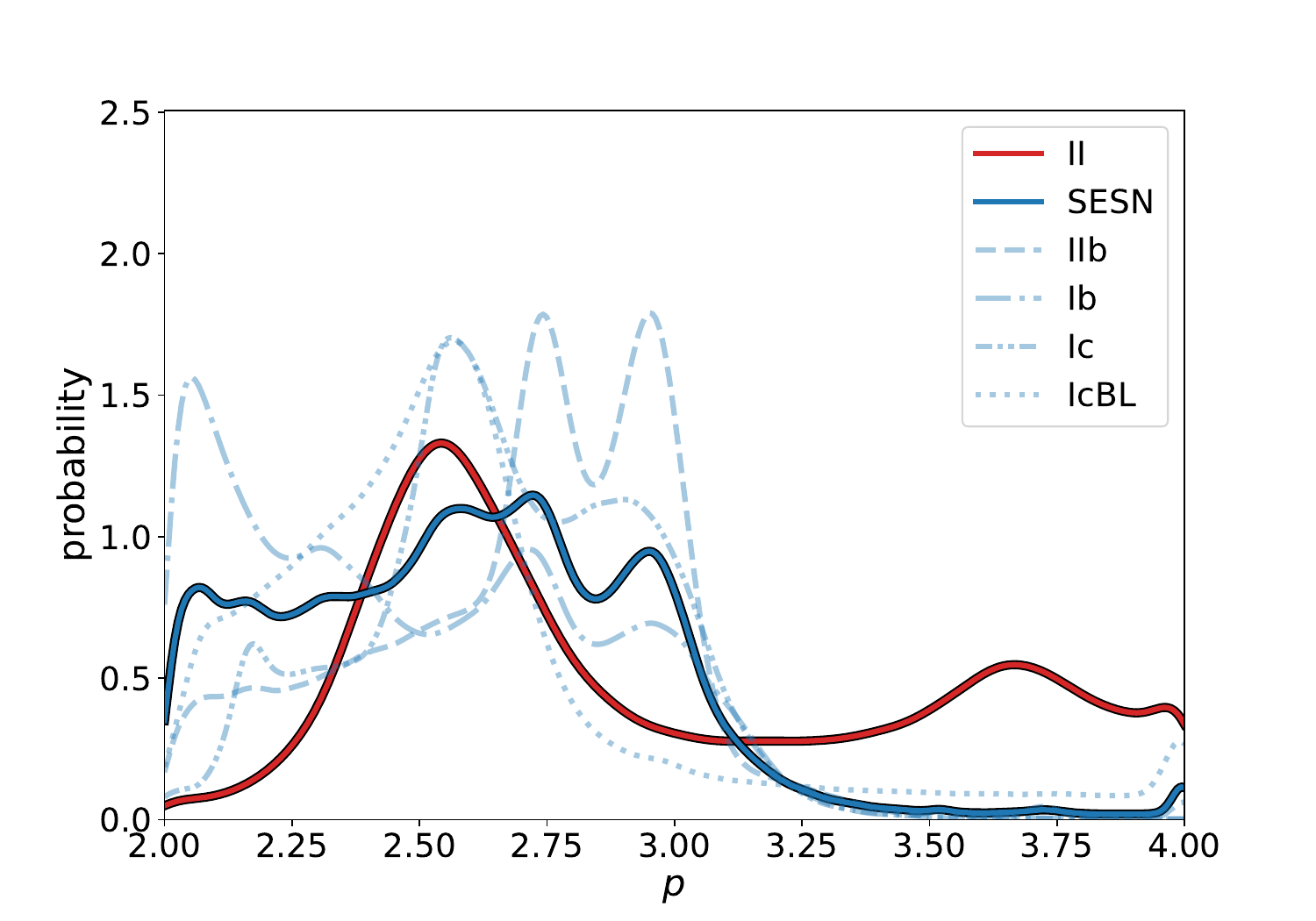} \\
\includegraphics[width=0.5\linewidth]{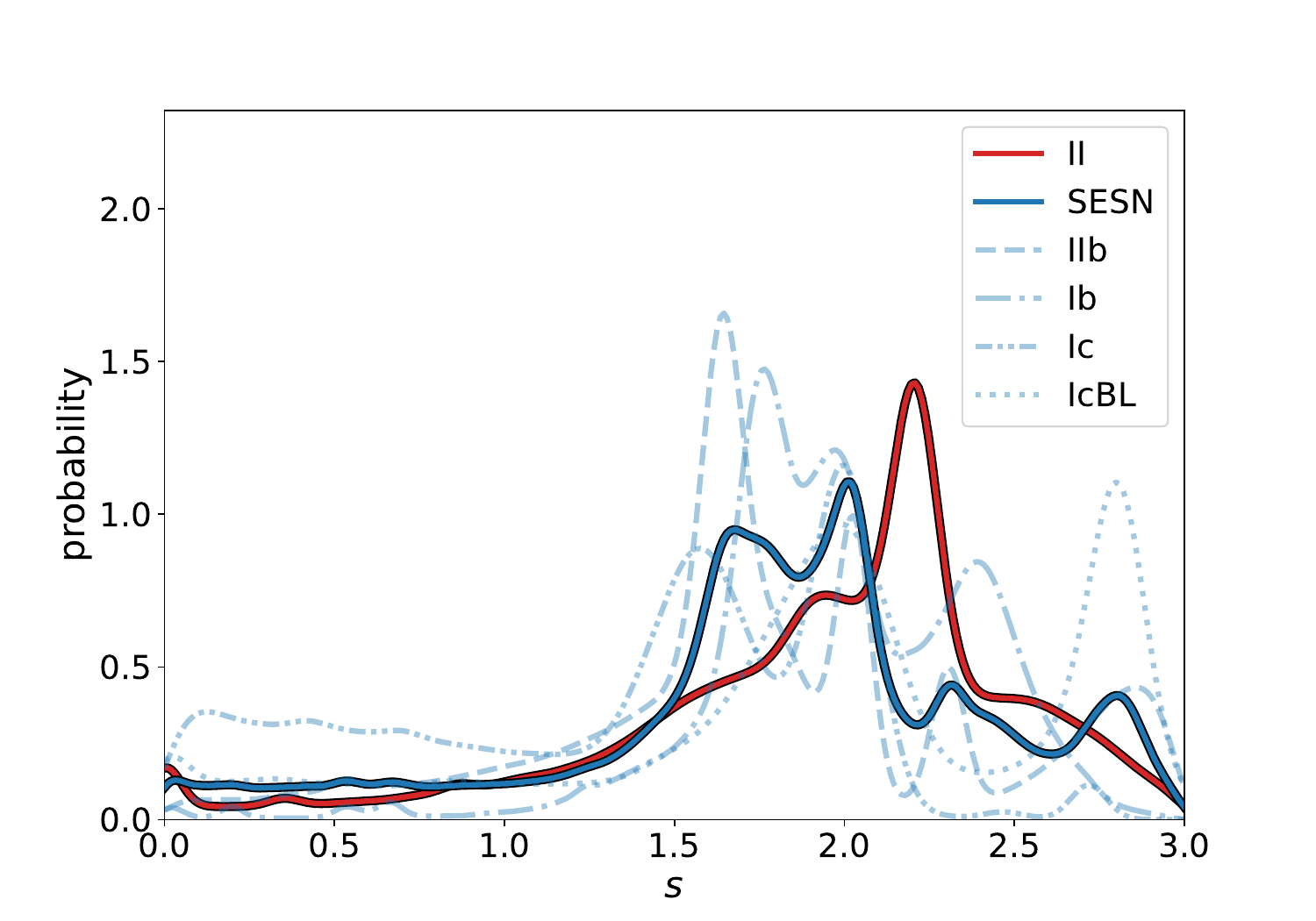}
\includegraphics[width=0.5\linewidth]{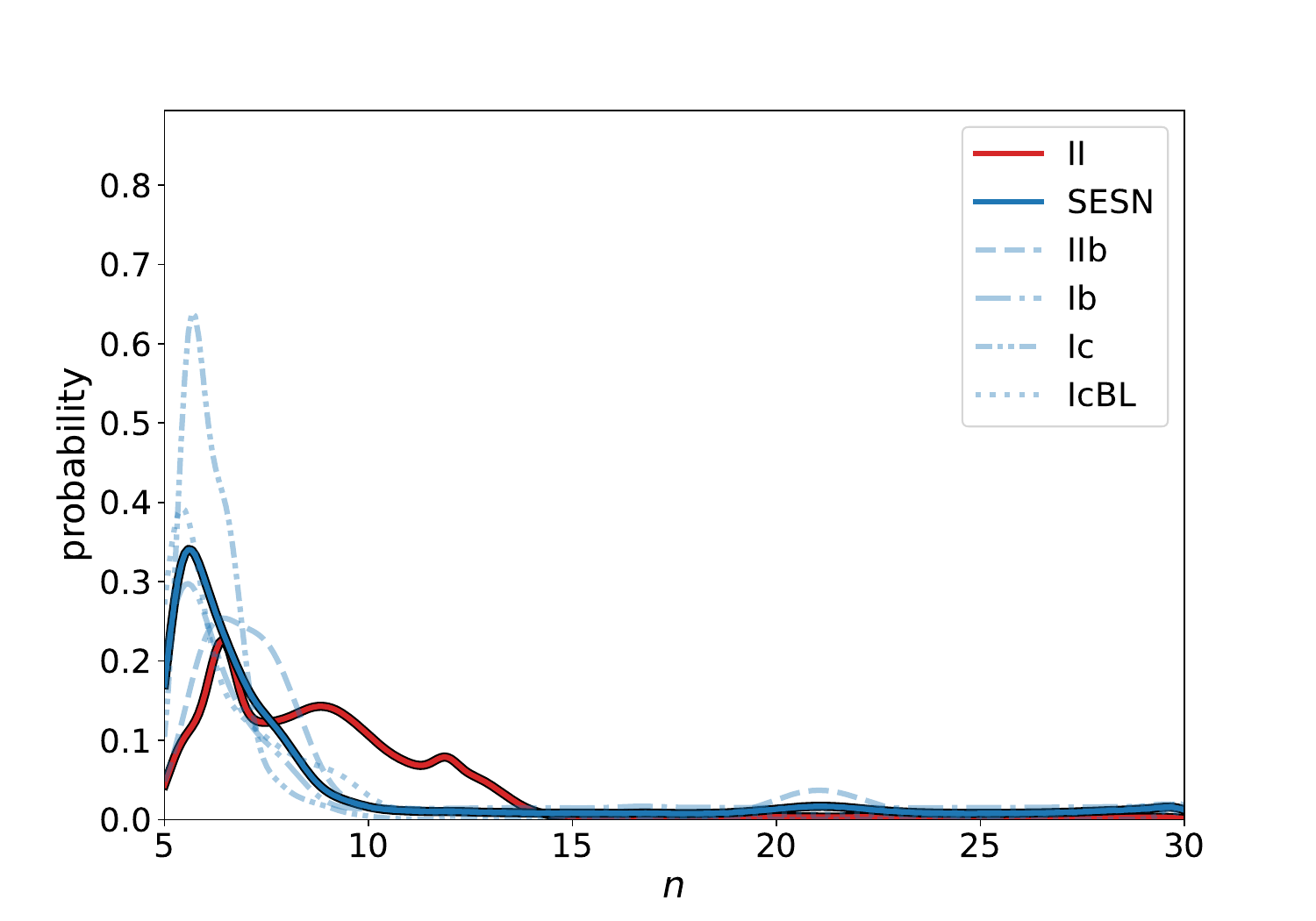}
\caption{The marginalized one-dimensional probability density functions of surveyed parameters: (a) CSM density scale $\logAast$, (b) $\logee$, (c) $\logeb$, (d) spectral index $p$, (e) CSM slope $s$, and (f) ejecta slope $n$.}\label{fig:PDF59}
\end{figure*}

We note that the credible intervals derived above may not necessarily cover all the fitting uncertainties. As an example, in Figure~1, we plot the model light curves based on the parameter sets with the 1st, 11th, 21st, ..., 91st smallest values of $\chi^2_{\rm red}$ in different colors, extracted from the 100 walkers applied to SN~2003L. We find that even the parameter set with the lowest grade provides at least a marginal fit to the observational data in terms of visual inspection, even if the parameter set is outside the credible interval. With this caveat in mind, we still employ the quantified credible interval as a useful indicator for the later discussion on the physical properties of radio SN parameters.

\begin{table*}[htbp]
    \centering
	\caption{Summary of the fitting results of $\logAast, \logee$, and $\logeb$ for successfully fitted radio SN samples.}\label{tab:median_1sigma_eachSN_logAlogeelogeB}
	\begin{tabular}{cccccccc} 
		\hline
		\hline
        & & mode of & median of & mode of & median of & mode of & median of \\
        SN type & SN name & $\logAast$ & $\logAast$ & $\logee$ & $\logee$ & $\logeb$ & $\logeb$ \\
		\hline
		II & SN 2004dj & $1.61$ & $2.07\pm_{0.68}^{1.55}$ & $-3.85$ & $-3.66\pm_{0.70}^{1.25}$ & $-0.25$ & $-0.97\pm_{2.86}^{0.70}$ \\
 & SN 2012aw & $1.29$ & $1.35\pm_{0.88}^{0.96}$ & $-0.45$ & $-1.56\pm_{1.77}^{1.04}$ & $-0.42$ & $-1.74\pm_{1.59}^{1.15}$ \\
 & SN 2016X & $2.66$ & $2.21\pm_{1.20}^{0.90}$ & $-2.12$ & $-2.26\pm_{1.64}^{1.44}$ & $-2.22$ & $-2.42\pm_{1.51}^{1.44}$ \\
\hline 
IIb & SN 2001gd & $2.54$ & $2.56\pm_{0.62}^{0.80}$ & $-3.53$ & $-2.16\pm_{1.24}^{1.15}$ & $-4.21$ & $-3.03\pm_{1.34}^{1.45}$ \\
 & SN 2001ig & $1.79$ & $1.73\pm_{0.62}^{0.60}$ & $-1.34$ & $-1.40\pm_{0.80}^{0.67}$ & $-0.99$ & $-1.40\pm_{1.14}^{0.85}$ \\
 & SN 2011dh & $1.39$ & $1.37\pm_{0.42}^{0.56}$ & $-2.51$ & $-2.84\pm_{1.04}^{0.90}$ & $-0.43$ & $-1.20\pm_{1.61}^{0.79}$ \\
 & SN 2011ei & $0.99$ & $0.87\pm_{0.94}^{0.84}$ & $-1.20$ & $-2.49\pm_{1.49}^{1.32}$ & $-0.27$ & $-2.49\pm_{1.67}^{1.74}$ \\
 & SN 2011hs & $1.23$ & $1.21\pm_{0.20}^{0.18}$ & $-2.94$ & $-2.88\pm_{0.30}^{0.40}$ & $-0.18$ & $-0.54\pm_{0.52}^{0.33}$ \\
 & SN 2013df & $3.56$ & $3.50\pm_{0.30}^{0.26}$ & $-3.24$ & $-3.23\pm_{0.28}^{0.30}$ & $-2.78$ & $-2.79\pm_{0.28}^{0.32}$ \\
 & SN 2016gkg & $1.97$ & $1.63\pm_{1.53}^{0.50}$ & $-0.45$ & $-0.57\pm_{0.52}^{0.17}$ & $-3.91$ & $-3.44\pm_{0.79}^{3.01}$ \\
\hline 
Ib & SN 1983N & $0.89$ & $0.81\pm_{1.00}^{0.96}$ & $-2.64$ & $-2.78\pm_{0.97}^{0.85}$ & $-0.23$ & $-1.49\pm_{1.34}^{1.00}$ \\
 & SN 2004dk & $0.61$ & $0.67\pm_{0.96}^{1.00}$ & $-2.47$ & $-2.34\pm_{0.77}^{0.85}$ & $-1.20$ & $-1.56\pm_{1.14}^{0.95}$ \\
 & SN 2004gq & $1.03$ & $1.03\pm_{0.20}^{0.36}$ & $-3.86$ & $-3.70\pm_{0.25}^{1.94}$ & $-0.12$ & $-0.30\pm_{3.39}^{0.20}$ \\
 & SN 2008D & $-0.88$ & $-0.70\pm_{0.66}^{0.98}$ & $-2.07$ & $-2.34\pm_{0.90}^{0.58}$ & $-0.27$ & $-1.51\pm_{1.96}^{1.07}$ \\
 & SN 2012au & $1.41$ & $1.45\pm_{0.58}^{0.68}$ & $-2.17$ & $-2.59\pm_{0.97}^{0.70}$ & $-1.49$ & $-2.22\pm_{1.59}^{1.45}$ \\
 & AT2014ge & $0.35$ & $0.27\pm_{0.30}^{0.32}$ & $-1.45$ & $-2.66\pm_{0.92}^{1.12}$ & $-0.08$ & $-0.30\pm_{1.34}^{0.20}$ \\
 & SN 2014C & $2.76$ & $2.54\pm_{0.64}^{0.54}$ & $-0.67$ & $-1.15\pm_{0.74}^{0.50}$ & $-3.93$ & $-3.83\pm_{0.64}^{0.74}$ \\
\hline 
Ic & SN 1990B & $-0.78$ & $-0.56\pm_{0.50}^{0.80}$ & $-3.61$ & $-3.63\pm_{0.89}^{0.97}$ & $-0.70$ & $-1.57\pm_{1.47}^{1.05}$ \\
 & SN 1994I & $2.49$ & $2.35\pm_{0.30}^{0.16}$ & $-1.00$ & $-1.07\pm_{0.25}^{0.17}$ & $-4.85$ & $-4.53\pm_{0.33}^{0.69}$ \\
 & SN 2003L & $1.31$ & $1.29\pm_{0.14}^{0.14}$ & $-2.54$ & $-2.37\pm_{0.23}^{0.52}$ & $-0.60$ & $-0.75\pm_{0.69}^{0.45}$ \\
 & SN 2004cc & $1.37$ & $1.45\pm_{0.38}^{1.79}$ & $-2.19$ & $-2.26\pm_{0.58}^{0.30}$ & $-0.47$ & $-0.80\pm_{2.21}^{0.48}$ \\
 & SN 2020oi & $1.63$ & $1.41\pm_{1.08}^{0.84}$ & $-1.91$ & $-2.27\pm_{1.42}^{0.84}$ & $-4.63$ & $-3.04\pm_{1.39}^{1.97}$ \\
\hline 
IcBL & SN 2003bg & $2.68$ & $2.66\pm_{0.12}^{0.12}$ & $-1.49$ & $-1.51\pm_{0.12}^{0.10}$ & $-1.91$ & $-1.92\pm_{0.33}^{0.33}$ \\
 & SN 2007bg & $1.87$ & $1.93\pm_{0.34}^{0.46}$ & $-1.94$ & $-1.96\pm_{0.42}^{0.54}$ & $-1.82$ & $-2.17\pm_{1.27}^{0.90}$ \\
 & SN 2009bb & $-0.54$ & $-0.54\pm_{0.20}^{0.24}$ & $-2.99$ & $-3.03\pm_{0.12}^{0.10}$ & $-0.07$ & $-0.13\pm_{0.17}^{0.08}$ \\
 & SN 2012ap & $0.61$ & $0.31\pm_{1.18}^{1.16}$ & $-2.29$ & $-2.56\pm_{0.99}^{0.74}$ & $-1.07$ & $-2.22\pm_{1.74}^{1.44}$ \\
 & SN 2016coi & $2.21$ & $1.87\pm_{0.74}^{0.60}$ & $-1.97$ & $-2.22\pm_{0.95}^{0.79}$ & $-1.51$ & $-1.74\pm_{1.27}^{1.02}$ \\
		\hline
		\hline
	\end{tabular}
\end{table*}

\begin{table*}[htbp]
    \centering
	\caption{Summary of the fitting results of $p, s$, and $n$ for successfully fitted radio SN samples.}\label{tab:median_1sigma_eachSN_psn}
	\begin{tabular}{cccccccc} 
		\hline
		\hline
        & & mode of & median of & mode of & median of & mode of & median of \\
        SN type & SN name & $p$ & $p$ & $s$ & $s$ & $n$ & $n$ \\
		\hline
		II & SN 2004dj & $2.51$ & $2.62\pm_{0.19}^{1.02}$ & $2.21$ & $2.14\pm_{0.54}^{0.10}$ & $6.42$ & $7.84\pm_{1.59}^{1.51}$ \\
 & SN 2012aw & $2.61$ & $2.59\pm_{0.19}^{0.20}$ & $1.97$ & $2.00\pm_{0.17}^{0.18}$ & $11.90$ & $10.50\pm_{2.68}^{2.01}$ \\
 & SN 2016X & $2.70$ & $2.80\pm_{0.33}^{0.57}$ & $2.46$ & $2.33\pm_{0.95}^{0.30}$ & $5.50$ & $7.09\pm_{1.67}^{2.68}$ \\
\hline 
IIb & SN 2001gd & $2.80$ & $2.79\pm_{0.34}^{0.32}$ & $1.53$ & $1.31\pm_{0.61}^{0.45}$ & $6.09$ & $6.00\pm_{0.42}^{0.50}$ \\
 & SN 2001ig & $2.10$ & $2.23\pm_{0.16}^{0.28}$ & $2.86$ & $2.76\pm_{0.23}^{0.13}$ & $5.67$ & $6.51\pm_{0.92}^{1.17}$ \\
 & SN 2011dh & $2.82$ & $2.74\pm_{0.23}^{0.12}$ & $1.78$ & $1.77\pm_{0.14}^{0.12}$ & $6.59$ & $6.59\pm_{0.58}^{0.75}$ \\
 & SN 2011ei & $2.59$ & $2.61\pm_{0.25}^{0.28}$ & $1.46$ & $1.27\pm_{0.71}^{0.28}$ & $7.17$ & $9.18\pm_{2.17}^{4.68}$ \\
 & SN 2011hs & $3.95$ & $3.86\pm_{0.16}^{0.09}$ & $1.25$ & $1.22\pm_{0.21}^{0.20}$ & $5.67$ & $5.67\pm_{0.25}^{0.33}$ \\
 & SN 2013df & $2.78$ & $2.77\pm_{0.18}^{0.17}$ & $2.02$ & $2.01\pm_{0.09}^{0.10}$ & $29.40$ & $28.20\pm_{2.42}^{1.17}$ \\
 & SN 2016gkg & $2.96$ & $2.94\pm_{0.06}^{0.05}$ & $1.65$ & $1.63\pm_{0.04}^{0.04}$ & $7.59$ & $7.68\pm_{0.42}^{0.58}$ \\
\hline 
Ib & SN 1983N & $2.97$ & $2.96\pm_{0.13}^{0.13}$ & $1.89$ & $1.72\pm_{0.31}^{0.28}$ & $5.17$ & $5.50\pm_{0.42}^{1.00}$ \\
 & SN 2004dk & $2.03$ & $2.07\pm_{0.05}^{0.11}$ & $2.29$ & $2.25\pm_{0.42}^{0.33}$ & $5.33$ & $5.67\pm_{0.50}^{1.17}$ \\
 & SN 2004gq & $2.34$ & $2.36\pm_{0.11}^{0.12}$ & $1.99$ & $1.96\pm_{0.12}^{0.09}$ & $5.50$ & $5.59\pm_{0.42}^{0.50}$ \\
 & SN 2008D & $2.88$ & $2.86\pm_{0.20}^{0.19}$ & $0.76$ & $0.70\pm_{0.39}^{0.38}$ & $6.34$ & $6.25\pm_{0.50}^{0.50}$ \\
 & SN 2012au & $2.85$ & $2.84\pm_{0.22}^{0.21}$ & $2.04$ & $1.86\pm_{0.41}^{0.30}$ & $5.17$ & $5.50\pm_{0.42}^{0.67}$ \\
 & AT2014ge & $2.72$ & $2.66\pm_{0.12}^{0.08}$ & $1.75$ & $1.74\pm_{0.06}^{0.06}$ & $6.92$ & $7.17\pm_{0.75}^{0.75}$ \\
 & SN 2014C & $2.33$ & $2.50\pm_{0.32}^{0.40}$ & $1.85$ & $1.84\pm_{0.14}^{0.13}$ & $28.80$ & $24.10\pm_{6.02}^{4.10}$ \\
\hline 
Ic & SN 1990B & $2.82$ & $2.81\pm_{0.19}^{0.18}$ & $0.12$ & $0.50\pm_{0.36}^{0.49}$ & $5.67$ & $5.75\pm_{0.33}^{0.50}$ \\
 & SN 1994I & $2.58$ & $2.58\pm_{0.09}^{0.10}$ & $2.01$ & $1.99\pm_{0.09}^{0.08}$ & $6.51$ & $6.42\pm_{0.25}^{0.42}$ \\
 & SN 2003L & $2.17$ & $2.34\pm_{0.17}^{0.19}$ & $1.57$ & $1.57\pm_{0.15}^{0.15}$ & $5.67$ & $5.75\pm_{0.50}^{0.75}$ \\
 & SN 2004cc & $2.80$ & $2.80\pm_{0.39}^{0.37}$ & $2.89$ & $2.81\pm_{2.46}^{0.12}$ & $5.25$ & $5.25\pm_{0.17}^{0.25}$ \\
 & SN 2020oi & $2.94$ & $2.92\pm_{0.19}^{0.15}$ & $1.72$ & $1.40\pm_{0.83}^{0.42}$ & $5.50$ & $6.17\pm_{0.84}^{1.67}$ \\
\hline 
IcBL & SN 2003bg & $2.60$ & $2.58\pm_{0.10}^{0.09}$ & $2.81$ & $2.78\pm_{0.07}^{0.07}$ & $5.33$ & $5.42\pm_{0.33}^{0.42}$ \\
 & SN 2007bg & $2.41$ & $2.40\pm_{0.23}^{0.23}$ & $1.91$ & $1.89\pm_{0.21}^{0.20}$ & $5.50$ & $5.59\pm_{0.42}^{0.75}$ \\
 & SN 2009bb & $2.96$ & $2.94\pm_{0.09}^{0.08}$ & $1.51$ & $1.49\pm_{0.12}^{0.12}$ & $5.67$ & $5.67\pm_{0.42}^{0.50}$ \\
 & SN 2012ap & $2.41$ & $2.47\pm_{0.27}^{0.35}$ & $0.28$ & $0.97\pm_{0.67}^{0.90}$ & $5.59$ & $6.34\pm_{0.92}^{1.17}$ \\
 & SN 2016coi & $2.41$ & $2.39\pm_{0.19}^{0.19}$ & $2.03$ & $1.98\pm_{0.19}^{0.14}$ & $7.17$ & $7.51\pm_{1.34}^{1.59}$ \\
		\hline
		\hline
	\end{tabular}
\end{table*}

Applying the same procedure to all radio SN samples enables us to obtain the 1D marginalized probability density functions of the parameters for each SN. Further, we construct the probability density functions of each parameter for each SN type, as illustrated in Figure~\ref{fig:PDF59}. Here, SNe IIb, Ib, Ic, and IcBL are grouped as the `SESN' type.
The modes and medians for each SN type are summarized in Table~\ref{tab:median_1sigma_SNtype_logAlogeelogeB} and \ref{tab:median_1sigma_SNtype_psn}.

Inspecting the distributions for each SN type allows us to discuss the nature of mass-loss activities of the progenitors and shock acceleration physics. $\logAast$ represents the CSM density scale and we can see that the median of $\logAast$ in SNe II is slightly larger than that of SESNe (but see Section~\ref{subsec:CSM_largescale}). Yet we note that the associated $1\sigma$ credible interval is large. 
Dividing the samples of SESNe into subclasses, we find that the median of $\logAast$ in SNe IcBL is the largest. Yet we note that this result is led by the fact that the probability density function of $\logAast$ in SN\,2003bg prefers the higher value, which drags the overall median of $\logAast$ towards the higher value. The median of $\logAast$ in SNe IIb, Ib, and Ic are $1.59$, $1.17$, and $1.33$, respectively. We can see $1\sigma$ credible intervals comparable to 1, indicating the presence of uncertainty by orders of magnitude.

Next, we discuss the medians of $\logee$ and $\logeb$, both of which are relevant to the nature of shock acceleration physics.
One of the frequent assumptions on $\epsilon_e$ and $\epsilon_B$ is to consider energy equipartition; $\epsilon_e=\epsilon_B$ with the absolute value depending on papers (\citealp[0.1 in][]{2013MNRAS.428.1207S} and 0.33 in \citealp[][]{2022ApJ...934..186N}).
It is also the case that $\epsilon_e=0.1$ and $\epsilon_B=0.01$ are ad hoc assumed to fit the observed radio light curves and discuss the nature of the CSM \citep{2019ApJ...883..147T,2023ApJ...951L..31B}.
Our parameter estimation indicates that the medians of $\logee$ in SN II and SESNe are $-3.09$ and $-2.11$, respectively, both of which are orders of magnitude lower than the abovementioned values. The medians of $\logeb$ in SNe II and SESNe are $-2.12$ and $-2.07$, respectively. This is more or less comparable with the assumption of $\epsilon_B=0.01$. However, we note that the values of mode and median of $\logeb$ vary substantially for different objects. The shapes of the probability density functions of $\logee$ and $\logeb$ are all roughly flat, involving uncertainty corresponding to an order of magnitude. This is found from the $1\sigma$ credible intervals of $\logee$ and $\logeb$ ranging from 1 to 2, implying the presence of uncertainty by orders of magnitude.

The probability density functions of the spectral index of relativistic electrons ($p$) are mainly characterized by the peak in the range of $2.5 < p < 3.0$. Even though the spectrum with this spectral index is softer than theoretical predictions from diffusive shock acceleration \citep[e.g.,][]{1978ApJ...221L..29B,1978MNRAS.182..147B,1983RPPh...46..973D,1987PhR...154....1B}, this tendency is consistent with previous indications from radio SN observations \citep[see e.g.,][and the references of Table\,\ref{tab:sample}]{1998ApJ...509..861F,2002ARA&A..40..387W,2006ApJ...641.1029C,2006ApJ...651..381C,2012ApJ...758...81M}.

The power-law index of CSM density profile ($s$) shows a potentially interesting tendency. 
Comparing the probability density functions of $s$ with the benchmark value of $s=2$ is important because it stands for steady wind distribution. $s>2$ describes the increasing mass-loss rate of the progenitor towards the core collapse, while $s<2$ indicates the decreasing mass-loss rate.
Our results are largely consistent with $s=2$ for almost all SN types. Yet, we find the modes and medians smaller than $s=2$ for SESNe (except for the IcBL class), implying the decreasing mass-loss rate towards core collapse. We will expand further discussion in Section~\ref{subsec:s_SESNe}.

However, we note that this trend of the probability density functions of $s$ may be potentially affected by the distribution of $n$, the density slope of the outer ejecta. Both $s$ and $n$ have large contributions to characterize the slope of the radio light curve in SNe \citep{2013ApJ...762...14M,2021ApJ...918...34M}. In fact, the probability density functions of $n$ in SESNe are concentrated on $n\sim5$, while that of SNe II has a peak at $n\sim7$. Note that $n=5$ is a critical value for describing the density structure of the ejecta \citep{1982ApJ...258..790C,1999ApJS..120..299T}. These values are smaller than the previous theoretical prediction \citep{1999ApJ...510..379M}, although the fact that the median of $n$ in SESNe looks smaller than that in SNe II is in line with the prediction. We will expand the detailed analysis and discussion in Section\,\ref{sec:Discussion_ejecta}.

\section{Discussion on the nature of CSM}\label{sec:Discussion_CSM}
\subsection{CSM structure in the outer region $r\gtrsim10^{15}\,{\rm cm}$}\label{subsec:CSM_largescale}

\begin{table*}[htbp]
	\centering
	\caption{Summary of the fitting results of $\logAast, \logee$, and $\logeb$ for each SN type.}\label{tab:median_1sigma_SNtype_logAlogeelogeB}
	\begin{tabular}{cccccccc} 
		\hline
		\hline
        & mode of & median of & mode of & median of & mode of & median of \\
        SN type & $\logAast$ & $\logAast$ & $\logee$ & $\logee$ & $\logeb$ & $\logeb$ \\
		\hline
		II & $1.65$ & $1.75\pm_{1.71}^{1.53}$ & $-4.03$ & $-3.09\pm_{1.30}^{2.01}$ & $-0.47$ & $-2.12\pm_{1.86}^{1.61}$ \\
SESN & $1.33$ & $1.47\pm_{1.20}^{1.08}$ & $-1.51$ & $-2.11\pm_{1.40}^{1.12}$ & $-0.12$ & $-2.07\pm_{2.06}^{1.51}$ \\
IIb & $1.45$ & $1.59\pm_{0.76}^{0.98}$ & $-0.50$ & $-1.47\pm_{1.72}^{1.04}$ & $-1.69$ & $-2.09\pm_{2.09}^{1.35}$ \\
Ib & $1.05$ & $1.17\pm_{0.94}^{1.36}$ & $-3.86$ & $-2.41\pm_{1.25}^{1.17}$ & $-0.12$ & $-2.11\pm_{1.94}^{1.79}$ \\
Ic & $1.31$ & $1.33\pm_{1.75}^{1.04}$ & $-1.02$ & $-2.31\pm_{1.40}^{1.22}$ & $-4.78$ & $-2.16\pm_{2.41}^{1.57}$ \\
IcBL & $2.68$ & $1.85\pm_{1.46}^{0.82}$ & $-1.51$ & $-2.07\pm_{1.27}^{0.62}$ & $-1.91$ & $-2.02\pm_{1.47}^{1.00}$\\
		\hline
		\hline
	\end{tabular}
\end{table*}

\begin{table*}[htbp]
	\centering
	\caption{Summary of the fitting results of $p, s$, and $n$ for each SN type.}\label{tab:median_1sigma_SNtype_psn}
	\begin{tabular}{ccccccc} 
		\hline
		\hline
        & mode of & median of & mode of & median of & mode of & median of \\
        SN type & $p$ & $p$ & $s$ & $s$ & $n$ & $n$ \\
		\hline
		II & $2.54$ & $2.74\pm_{0.29}^{0.90}$ & $2.21$ & $2.05\pm_{0.63}^{0.37}$ & $6.42$ & $8.43\pm_{2.26}^{3.34}$ \\
SESN & $2.72$ & $2.60\pm_{0.39}^{0.35}$ & $2.02$ & $1.85\pm_{0.55}^{0.55}$ & $5.59$ & $6.67\pm_{1.17}^{8.95}$ \\
IIb & $2.95$ & $2.74\pm_{0.40}^{0.23}$ & $1.65$ & $1.71\pm_{0.46}^{0.65}$ & $6.42$ & $7.34\pm_{1.34}^{12.50}$ \\
Ib & $2.05$ & $2.46\pm_{0.35}^{0.46}$ & $1.77$ & $1.95\pm_{0.27}^{0.44}$ & $5.59$ & $6.92\pm_{1.42}^{13.50}$ \\
Ic & $2.56$ & $2.65\pm_{0.27}^{0.30}$ & $2.00$ & $1.54\pm_{1.04}^{0.45}$ & $5.67$ & $6.00\pm_{0.58}^{0.84}$ \\
IcBL & $2.57$ & $2.52\pm_{0.29}^{0.31}$ & $2.80$ & $1.99\pm_{0.77}^{0.77}$ & $5.42$ & $6.51\pm_{1.17}^{4.77}$\\
		\hline
		\hline
	\end{tabular}
\end{table*}

\begin{figure}[htbp]
\includegraphics[width=\linewidth]{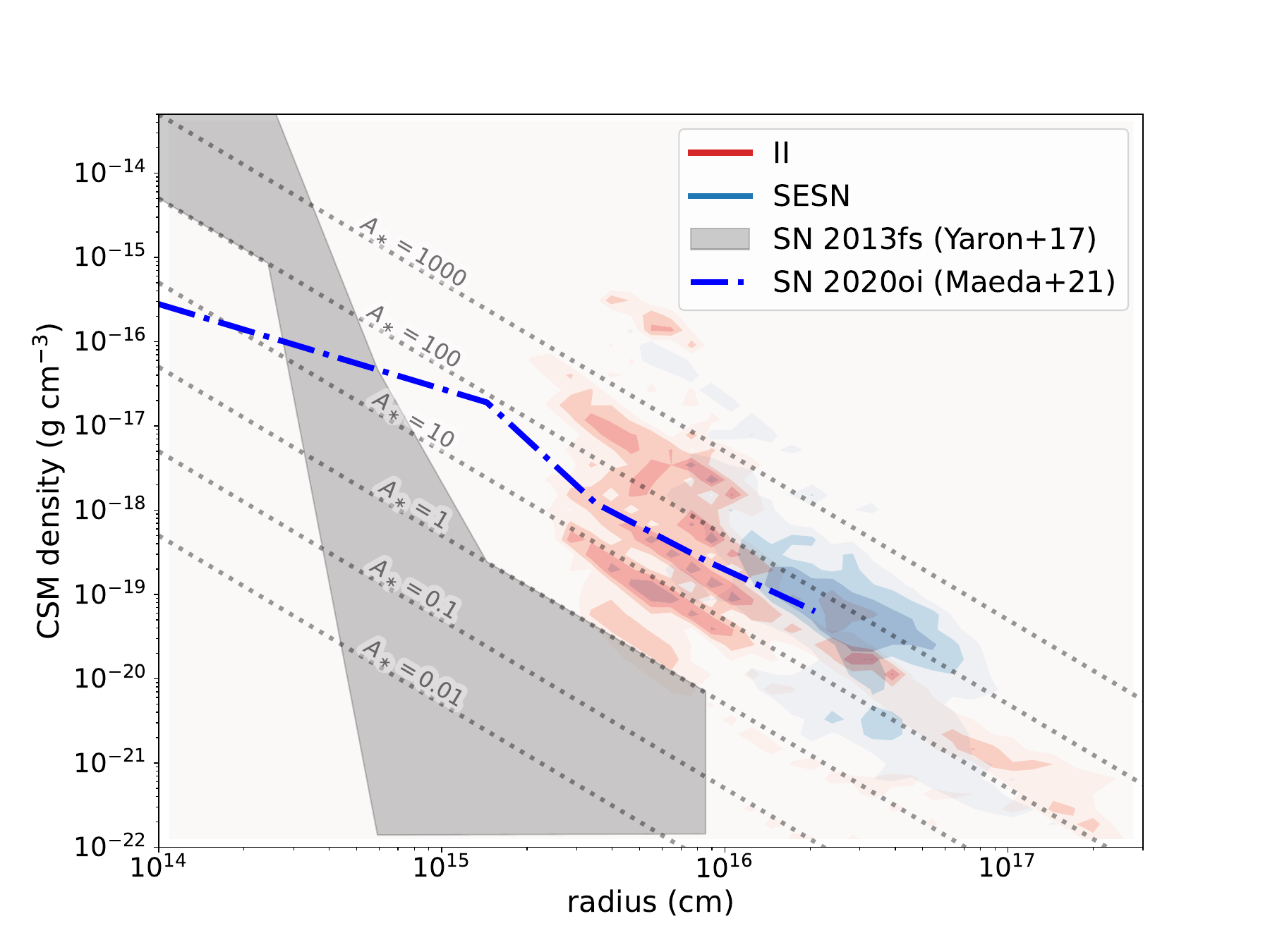}
\caption{The contour plot of CSM density profiles of SNe II (red) and SESNe (blue). The gray dotted lines represent steady wind profiles ($s=2$) with corresponding CSM density scales denoted ($A_\ast=\tilde{A}_\ast(s=2)$). For comparison we indicate the CSM models inferred for SN~2013fs by \citet{2017NatPh..13..510Y} and SN~2020oi by \citet{2021ApJ...918...34M}.}\label{fig:CSMcollection}
\end{figure}

Combining $\logAast$ and $s$ with Equation (\ref{eq:rhocsm}) enables us to construct CSM density profiles for each MCMC walker. Figure\,\ref{fig:CSMcollection} shows the contour plot of CSM density profiles divided into SNe II and SESNe.
Note that the inner and outer edges of the CSM density profiles are determined by the shock radius at the initial $(t_{\rm ini})$ and final observational time $(t_{\rm fin})$; the CSM distribution is limited to the radial range of $R_{\rm sh}(t_{\rm ini}) < r < R_{\rm sh}(t_{\rm fin})$. Since the data in the first 30~days are removed, the CSM density profiles are limited in the outer region of $r\gtrsim 3\times10^{15}\,{\rm cm}$. We also indicate the CSM models inferred for SN~2013fs \citep{2017NatPh..13..510Y} and SN~2020oi \citep{2021ApJ...918...34M} for reference.

It is found that CSM density profiles in SNe II exhibit a large scatter in the density scale ranging in $1\lesssim {A}_\ast\lesssim 1000$. This is actually reflecting the parameter estimation results obtained for individual SN II objects. The inner edge radius of the SN II contour is determined by the typical shock velocity of $10^4\,{\rm km\ s}^{-1}$, resulting in $30\,{\rm days}\times 10^9\,{\rm cm\ s}^{-1}\sim 3\times 10^{15}\,{\rm cm}$. It would be implied that the CSM density scale in SNe II has diverse values, while it is hard to identify systematic properties of $\tilde{A}_\ast$ in SNe II.

It is also found that the CSM density scale in SESNe is clustering at $A_\ast\sim100$. The shock velocity of SESNe is typically faster than that of SNe II due to smaller ejecta mass and sometimes higher kinetic energy. Thus the CSM radial length scale traced by radio SESNe is larger than the case of SNe II.

\begin{figure}[htbp]
\includegraphics[width=\linewidth]{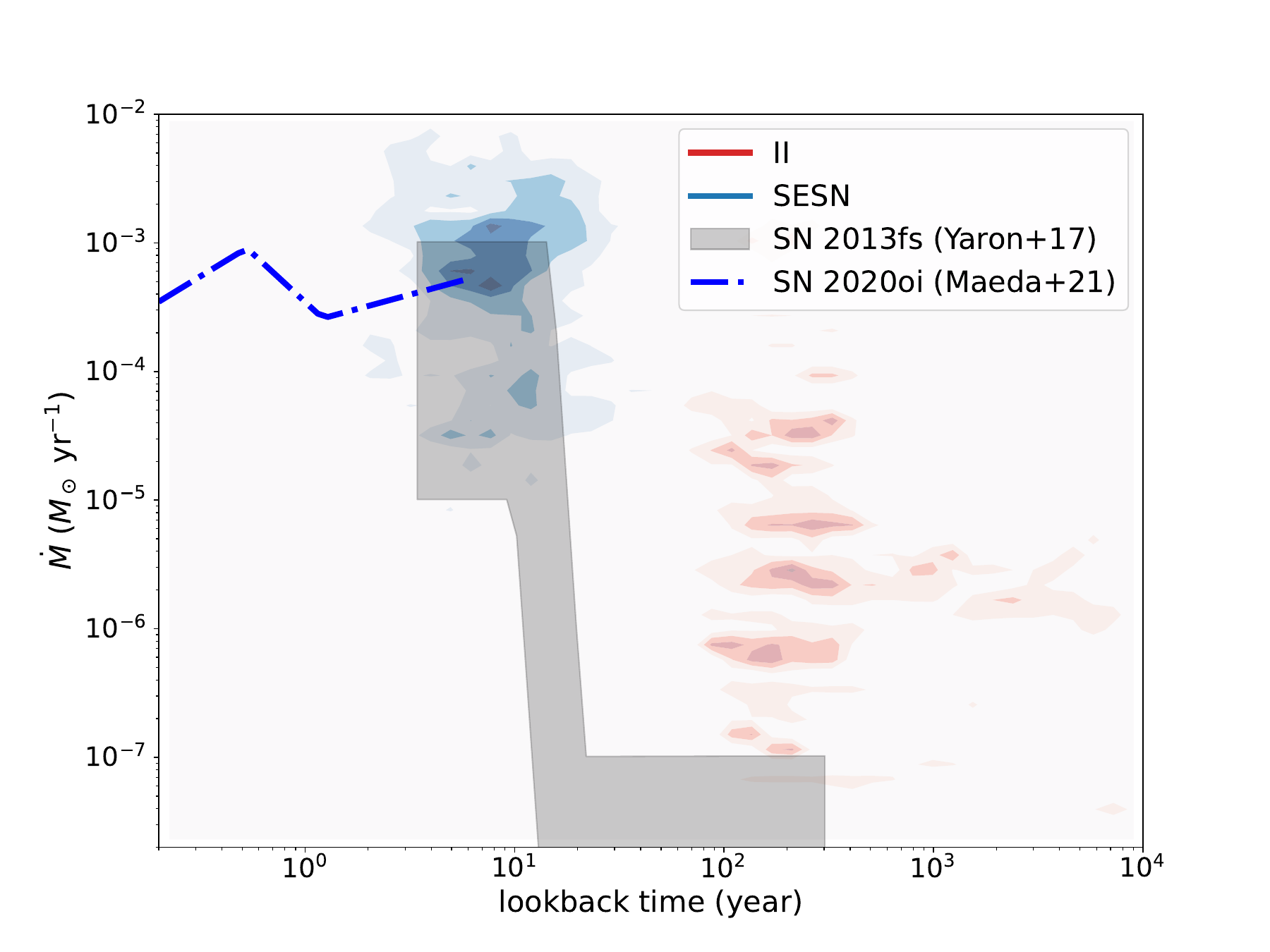}
\caption{The contour plot of mass-loss history models obtained from our parameter estimation. Red and blue contours represent the models of SN\,II and SESNe, respectively. The mass-loss history models for SN~2013fs \citep{2017NatPh..13..510Y} and SN~2020oi \citep{2021ApJ...918...34M} are also indicated.}\label{fig:mdot}
\end{figure}

Imposing the assumption on the velocity of the CSM allows us to visualize the time evolution of the mass-loss rate of the progenitors. Figure\,\ref{fig:mdot} shows the contour plot of mass-loss histories of SN progenitors in our sample. Here we assume that the CSM velocity is constant with time until the core collapse, and is comparable to the escape velocity of the progenitor; $\sim10\,{\rm km\ s^{-1}}$ for SNe II and $\sim1000\,{\rm km\ s^{-1}}$ for SESNe\footnote{Note that some progenitors of SNe IIb are inflated as much as yellow supergiants \citep{1994AJ....107..662A,2015ApJ...807...35M}, leading to the smaller CSM velocity than assumed.}.
We find that while the mass-loss rate of SNe II progenitors ranges by orders of magnitudes, the mass-loss rate of SESNe is clustering at the value of $\dot{M}\simeq 10^{-3}\,M_\odot{\rm yr}^{-1}$. This clear difference is attributed to the difference in the CSM density structure inferred in Figure\,\ref{fig:CSMcollection}.

The choice of the CSM velocity involves uncertainty and affects the location at which the mass-loss history model is plotted. The mass-loss rate derived here is proportional to the CSM velocity, while the look-back time on the abscissa is proportional to the inverse of the CSM velocity. Thus, the increase in the wind velocity leads to the shift of the mass-loss history model towards the upper left direction.
Still, it may be worth mentioning that the mass-loss history derived from our parameter estimation can be classified into two categories; one is those of SNe\,II with a range of mass-loss rate in the past ($\sim100\,$years before the explosion), and the other is those of SESNe with relatively higher mass-loss rate immediately ($\sim10\,$years) before the explosion.
The primitive interpretation is that we are observing the different mass-loss mechanisms between progenitors of SNe\,II and SESNe due to the drastic difference in the surface temperature of the progenitor stars. We also mention the possibility that this mass-loss rate evolution can be actually described by a unified model in which the mass-loss rate and the nuclear burning activity are interlocking with each other to get stronger and stronger, as advocated in the case study of SN\,2020oi \citep{2021ApJ...918...34M}.

It is worth mentioning that our estimate of $\dot{M}$ is by an order of magnitude larger than the result of \citet{2021ApJ...908...75B}. We suppose that the difference in the estimate is mainly attributed to the quantification of $\epsilon_e$ and $\epsilon_B$. \citet{2021ApJ...908...75B} assumed the equipartition of energy density between relativistic electrons and magnetic field and adopted $\epsilon_e=\epsilon_B=0.1$, the combination which leads to the modest CSM density scale. On the other hand, our parameter estimation includes $\epsilon_e$ and $\epsilon_B$ as fitting parameters and results in the preference of much lower values ($10^{-3}\lesssim \epsilon_e \lesssim 10^{-2}$ and$\epsilon_B \sim 0.01$).

Our SN samples are limited to those who have radio detections capturing a clear peak in their light curves, while more than half of SNe observed ever exhibit weak radio emission \citep{2021ApJ...908...75B} so that they elude detections by previous observations. Our analysis can thus be biased towards the SN events with bright radio emission, which prefers dense CSM, leading to a potential bias of picking up a sample of SNe with a large density scale ($\tilde{A}_\ast$) in our analysis. To eliminate this bias, deeper surveys of radio SNe with future facilities such as next-generation Very Large Array \citep{2018ASPC..517....3M} would be essential to accumulate detection data of radio-faint SN samples.

\subsection{Inferring the CSM structure in the vicinity of SN progenitors ($r\lesssim 10^{15}\,{\rm cm}$)}\label{subsec:CSM_vicinity}
\begin{figure}
    \centering
    \includegraphics[width=\linewidth]{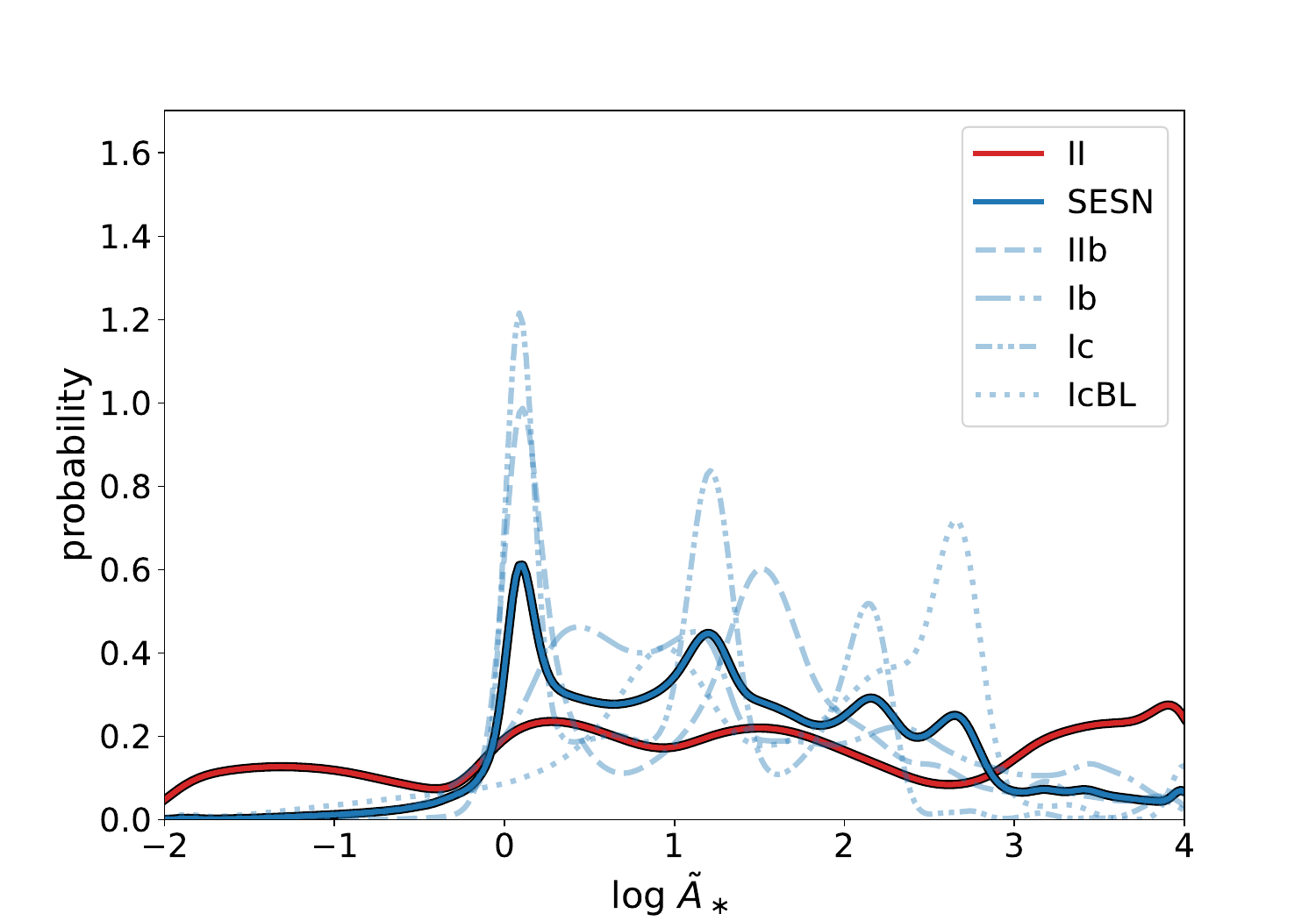}
    \includegraphics[width=\linewidth]{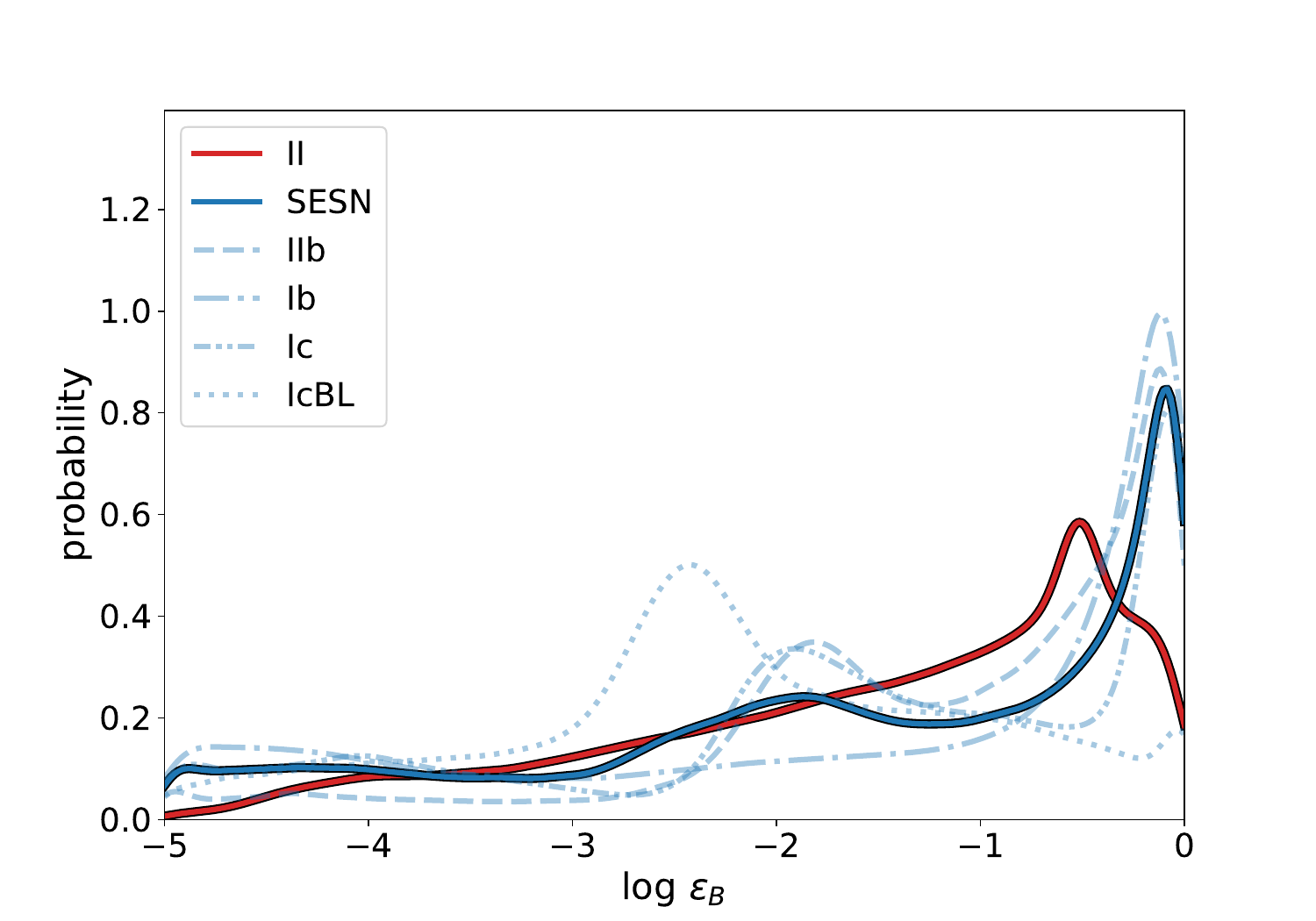}
    \caption{The marginalized one-dimensional probability density functions of (a) $\logAast$ and (b) $\logeb$ constructed from the MCMC sample data with the early-phase observational data included.}
    \label{fig:PDF_logA_logeB_early}
\end{figure}

In Section\,\ref{sec:fitting} we have shown our parameter estimation result where the early-phase ($< 30$ days) observational data are excluded (see Section\,\ref{sec:data_removal}).\footnote{Here we define ``early-phase" and ``late-phase" as the time ranges before and after 30~days since the explosion, respectively.} Actually, we also have conducted the parameter estimation with the early-phase data included, whose resultant probability density functions of $\logAast$ and $\logeb$ are displayed in Figure\,\ref{fig:PDF_logA_logeB_early}. We find that in this case, $\epsilon_B$ tends to be as large as $\epsilon_B \sim 1$, which is not achieved by current particle-in-cell simulations \citep[e.g.,][]{2011ApJ...726...75S, 2014ApJ...794...46C}. Instead, it is also found that the peak value of $\logAast$ is lower compared to the parameter estimation results with the early-phase data excluded (see Figure~\ref{fig:PDF59}). The probability density functions of the other parameters are displayed in Appendix\,\ref{app:allepoch}.

A straightforward interpretation is that this consequence is misled by the inclusion of the early-phase data, which implies that the behavior of the radio luminosity in the early phase is beyond the description of our current radio SN model.
The early-phase data removed in Section\,\ref{sec:fitting} includes (i) optically thick emission in a wide range of frequencies, or (ii) optically thin emission at high frequencies.
Taking this observational fact into account, it is required to find the parameter in $\vec{\theta}$ responsible for the consequence.

We guess that $s$ and $n$ are not relevant to this issue. These parameters are mainly determined by the rise and decline rates of radio light curves, which are well constrained by the data in the optically thin phase.
$p$ is not likely relevant to this issue either, because it is mainly related to the frequency dependence of the synchrotron spectrum at a given time. Thus, we suppose that the rest of the parameters, $\tilde{A}_\ast$ and $\epsilon_e$ behave indeed in the other way from our (fiducial) model and are responsible for producing the different consequence between two MCMC analyses. Hereafter we describe the response of the light curve to the variation of $\tilde{A}_\ast$ and $\epsilon_e$. We also refer readers to Appendix\,\ref{app:dependence} to figure out the general dependence of the radio light curve on the parameters under survey.
\begin{enumerate}
    \item If the removed data mainly contains optically thick emission, it means that the high $\epsilon_B$ is playing a role in decreasing the radio luminosity in the optically thick regime. This corresponds to the situation where the observed flux in the early phase is fainter than the extrapolation from the model consistent with observational data in the later phase. If we attempt to fit these fainter signals ``than expected" without enlarging $\epsilon_B$, we have to increase the normalization of the CSM density $\tilde{A}_\ast$ only in the inner region than the density in the outer region. The initially high $\epsilon_e$ cannot stand for the cause of high $\epsilon_B$ in this case because the radio luminosity of optically thick emission is independent of $\epsilon_e$.
    \item On the other hand, if the removed data mainly consists of optically thin emission in the high-frequency range, then it corresponds to the situation where the high $\epsilon_B$ was making the optically thin radio luminosity in the early phase brighter ``than expected" from the extrapolation from the later phase. To enhance the optically thin radio luminosity only in the early phase, we need to get either higher $\tilde{A}_\ast$ in the inner region than the outer CSM, or the greater $\epsilon_e$ in the early phase than in the late phase. 
\end{enumerate}

Based on these statements, we propose that the enhancement of the CSM density scale $\tilde{A}_\ast$ in the vicinity of the progenitor could solve the issue without introducing extremely high $\epsilon_B$. 
Data removal is forcing the parameter estimation procedure to focus on fitting only the late-phase data, where relatively low $\epsilon_B$ is found. To compensate for the discrepancy between the extrapolation of this model to the early phase and the observed data there, $\tilde{A}_\ast$ `only in the inner CSM' should generally be elevated (see Figure\,\ref{fig:PDF59}). This can be understood from the dependence of radio light curves on $\tilde{A}_\ast$ and $\epsilon_B$ being similar to each other (see Appendix\,\ref{app:dependence}). Interestingly, the presence of such a density enhancement has been indicated in the recent rapid follow-up optical observations of SNe, as mentioned in Section\,\ref{sec:data_removal}. From this perspective, it can be claimed that the extremely high $\epsilon_B$ implied from our parameter estimation could serve as indirect evidence of the enhanced mass-loss activity immediately before the explosion, irrespective of SN types. In addition to this, the initially higher $\epsilon_e$ could also explain the gap between parameter estimation with and without the early-phase data, as long as the removed data mainly includes optically thin emission. Yet, further examination should be required including multi-zone modeling of radio SNe or introducing time-dependent microphysics parameters to make these suggestions more robust \citep[e.g.,][]{2019ApJ...885...41M}.

\subsection{Shallower CSM density structure in SESNe}\label{subsec:s_SESNe}

In this subsection, we mention the shallow density slope of the CSM seen in SESNe. As shown in Table\,\ref{tab:median_1sigma_SNtype_psn}, the median value of $s$ in SESNe is calculated as 1.85. Although this value is consistent with that of SNe~II within the  $1\sigma$ credible interval, it is found that medians of $s$ in many samples of SESNe are smaller than $2$, indicating the density structure of CSM shallower than steady wind. Particularly, the mode and the median of $s$ in SNe IIb are both around $1.7$, implying a clear deviation from $s=2$. We can see that either the mode or the median of SNe~Ib and SNe~Ic is modestly smaller than $2$. This may suggest the possibility that the mass-loss activity of SESN progenitors is distinct from the progenitors of the other SN types. The value of $s$ smaller than 2 indicates the declining mass-loss rate towards the core collapse. It is still under debate what the exact mass-loss mechanism of SESN progenitors is. The advocated scenario includes binary interaction, single star's activity, and both of them \citep[see e.g.,][]{2019NatAs...3..434F}. Our results may serve as a key to understanding the nature of hydrogen-deficient stars' evolution.
Nevertheless, we note that this result is only indicative and needs further investigation, in view of the inconsistency between the values of mode and median seen in some columns.

\section{Discussion on shock acceleration physics}\label{sec:Discussion_PAMFA}

\subsection{The Validity of Equipartition between $\epsilon_e$ and $\epsilon_B$}\label{subsec:ee_eb}
It has been often assumed in the literature that the thermal energy dissipated from the SN shock will be distributed into relativistic electrons and magnetic field in the equipartition way. This situation is formalized by $\epsilon_e = \epsilon_B$. Their normalization is typically assumed to be orders of $\sim 0.1$ in fitting radio light curve of SNe \citep[see e.g.,][and references in Table \ref{tab:sample}]{2013MNRAS.428.1207S,2022ApJ...934..186N}. However, the real values of $\epsilon_e$ and $\epsilon_B$ are not yet clarified \citep[e.g.,][]{2006ApJ...641.1029C}. 
Not knowing the fiducial values of $\epsilon_e$ and $\epsilon_B$, the radio SN modeling can lead to a misleading result in the estimate of other parameters such as CSM density.

\begin{figure}
    \centering
    \includegraphics[width=\linewidth]{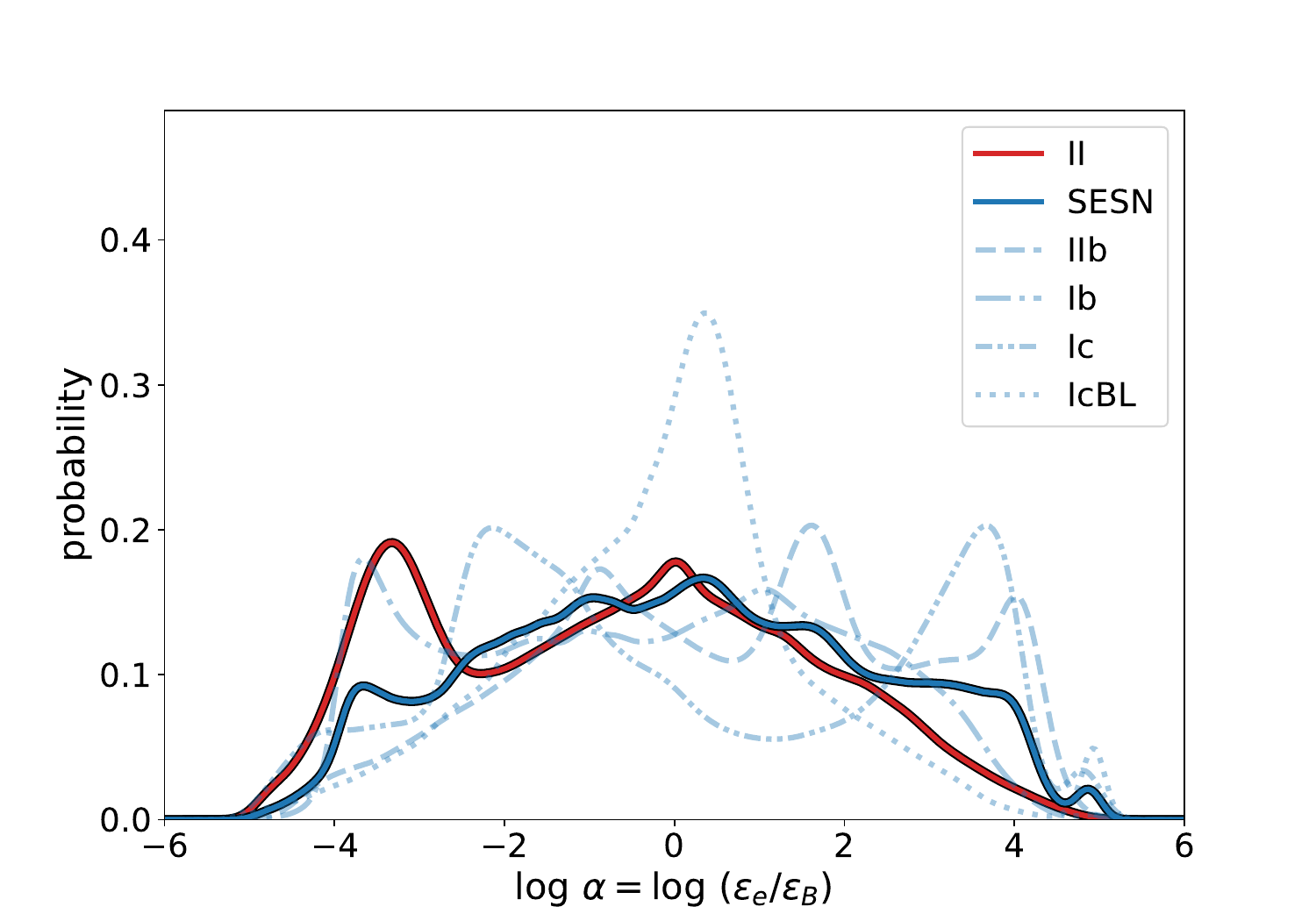}
    \includegraphics[width=\linewidth]{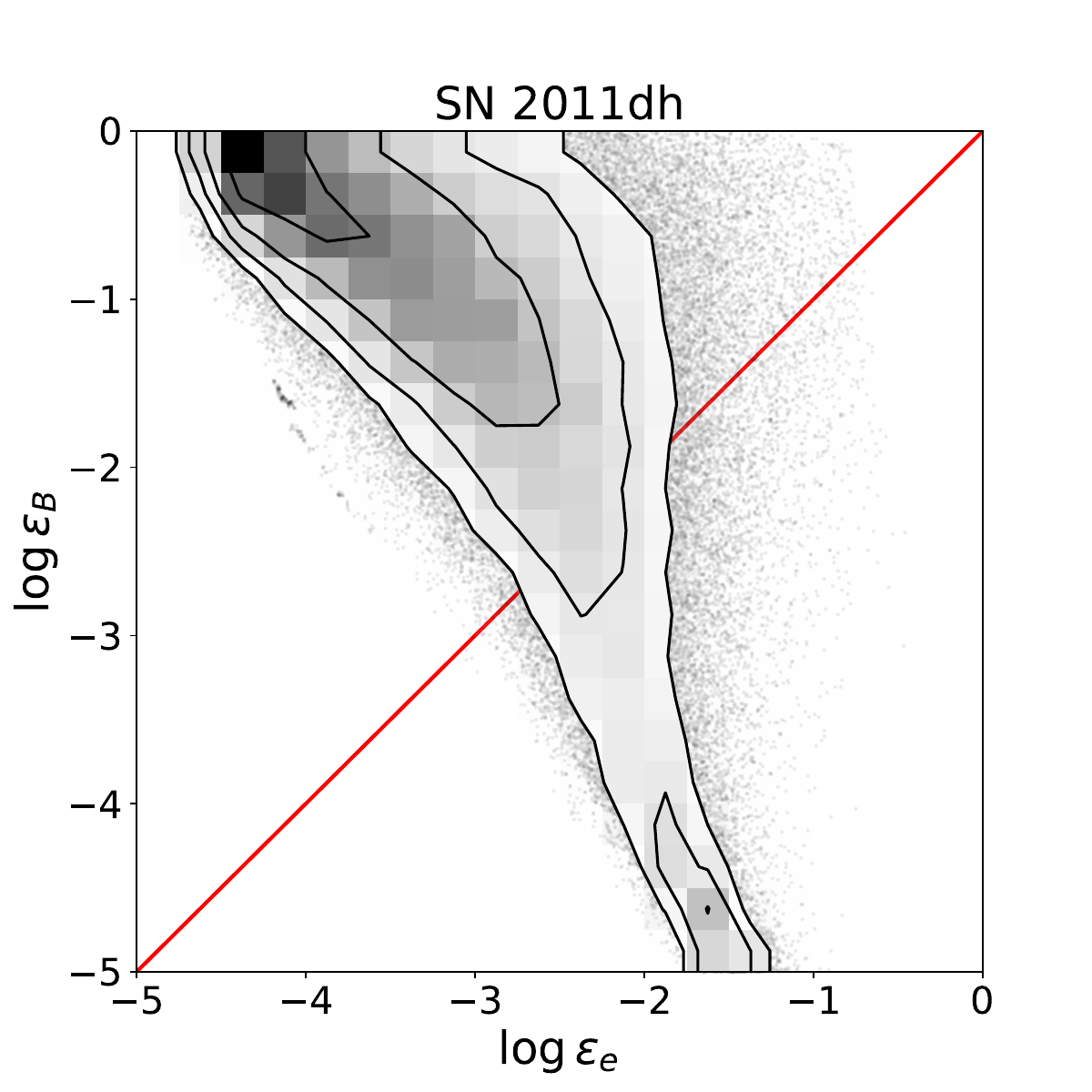}
    \caption{Top: The marginalized 1D probability density functions of $\log\alpha=\log(\epsilon_e/\epsilon_B)$. Bottom: The posterior distribution as functions of $\logee$ and $\logeb$ in SN~2011dh. The red line denotes the situation of equipartition ($\epsilon_e = \epsilon_B$).
    }
    \label{fig:epsilon}
\end{figure}

The top panel of Figure\,\ref{fig:epsilon} shows the probability density function of $\log\,\alpha = \log\,(\epsilon_e/\epsilon_B)$ for each SN type. The overall shape of the probability density distributions for SNe II and SESNe orient towards the value of $\log\alpha\simeq0$, indicating the equipartition of the energy density between electrons and magnetic field.
However, the posterior distribution of individual radio SN samples tends to cluster at the value deviating from equipartition, as shown in the case of SN~2011dh (see the bottom panel of Figure~\ref{fig:epsilon}), though the distribution is overlapping the line of $\epsilon_e = \epsilon_B$.
It is also found that the probability density functions of $\log\alpha$ in SESN subclasses do not necessarily take peaks at $\log\alpha\simeq0$. Thus, it can be said that while the overall distribution exhibits a peak centered at $\log\alpha\simeq0$, there are diverse shapes of functions depending on SN types, warranting careful interpretation of this result.
We note that there are several detailed behaviors seen in the probability density functions caused by the individual behaviors from each SN or subclass.

It is also found that the parameter space of inefficient shock acceleration (i.e., $\logee<-3$ ``and'' $\logeb<-3$) is not supported by any radio SN samples.
However, this tendency may be biased by the fact that faint radio SNe with lower $\epsilon_e$ and $\epsilon_B$ can easily elude observations (see also Section~\ref{subsec:origin-of-high-eb}).

\subsection{Origin of high $\epsilon_B$}\label{subsec:origin-of-high-eb}

In Section\,\ref{subsec:CSM_vicinity} we showed the interpretation that the inclusion of the early-phase data induces the misleading result of the high $\epsilon_B$. Yet, we cannot completely reject the possibility that this high $\epsilon_B$ is truly realized. There are some physical processes that reinforce the further amplification of the magnetic field in the shocked CSM. One possibility is the excitation of the turbulence in the CSM. Once an episode of mass eruption from the SN progenitor has taken place forming confined CSM, the erupted material plunges into the pre-existing CSM \citep[see e.g.,][]{2020A&A...635A.127K}, and it has potential to induce Rayleigh-Taylor instability to develop aspherically disordered hydrodynamical structure \citep{2025arXiv250314438H}. This will serve as a final configuration of the CSM, which will be hit by the ejecta of the subsequent SN. As a result, we can have a situation advantageous for the magnetic field amplification due to the disordered background hydrodynamical structure \citep{2009ApJ...695..825I,2022ApJ...936L...9T}. as well as a potential influence on electron acceleration \citep{2022ApJ...927..132A}.

Another possibility is the strong stellar magnetic field embedded in the background. In the stellar wind dominated by magnetohydrodynamics, a non-radial component of the magnetic field of the stellar wind scales with $\propto r^{-1}$ \citep{2022PhRvD.106l3025K}. Then the plasma beta $\beta=P_{\rm gas}/P_{\rm mag}$ becomes a decreasing function of radius, indicating that the circumstellar environment far from the progenitor could be highly magnetized. In this case, the SN shock propagates in the strongly magnetized CSM, and thus magnetic field could be strongly amplified. This scenario may be preferred in the case where the magnetic field inherent to the SN progenitor is strong, which has recently been reported in some helium stars \citep{2020MNRAS.499L.116H,2023Sci...381..761S}.

The absence of small $\epsilon_B$ seen in Figure~\ref{fig:epsilon} is the opposite tendency to the implications in the literature of GRB afterglows, where $10^{-5} \lesssim \epsilon_B \lesssim 10^{-2}$ has been proposed \citep[see e.g.,][]{2014ApJ...785...29S,2014MNRAS.442.3147B}. This may arise from the difference in nature between radio SNe and GRB afterglows. For instance, the SN shocks are nonrelativistic while the GRB jet propagates with the speed of a relativistic regime, probably resulting in different magnetic field amplifications. The difference in circumstellar environment configuration or the significance of turbulence could be also a cause for the different tendencies of $\epsilon_B$.

\subsection{Origin of the soft electron spectrum}\label{subsec:p_ind}
As discussed in Section\,\ref{sec:fitting}, the spectral index of the relativistic electrons prefers the soft value; $p>2.5$ for a large number of radio SN samples. This tendency is actually in agreement with the previous studies of radio SN modeling \citep[e.g.,][]{2006ApJ...641.1029C}. The suggested explanations for this soft spectrum include the non-linear effect at the particle acceleration \citep{2010APh....33..307C,2010MNRAS.407.1773C,2020ApJ...905....2C,2012ApJ...750..156L,2019ApJ...876...27Y}. It is also pointed out that the energy of electrons responsible for radio emission from SNe is not so energetic that they are out of the involvement of the diffusive shock acceleration. This implies that it is not an unnatural situation that the spectra of radio SNe are not compatible with the prediction of pure diffusive shock acceleration \citep{2013ApJ...762L..24M}.
Further, it is believed that a shock whose normal is perpendicular to the background magnetic field produces a soft particle spectrum \citep{2022ApJ...925...48X}. In cases where the toroidal component of the background magnetic field is prominent as discussed in Section\,\ref{subsec:origin-of-high-eb}, the SN shock can be naturally a perpendicular shock which may lead to the production of the soft spectra of relativistic particles.

\section{Discussion on SN ejecta density profile}\label{sec:Discussion_ejecta}

It should be desirable to discuss the origin of the preference of $n\sim5$ from our parameter estimation since this has not been proposed in the previous fitting of SNe II and SESNe except for relativistic SNe represented by SN~2012ap \citep{2014ApJ...797..107M}. 
It is not necessarily inconsistent with the previous studies, since the value of $n$ has often been fixed based on theoretical expectation \citep{1999ApJ...510..379M}; the possible deviation of $n$ from the theoretically expected value has been largely unexplored.
With smaller $n$, the decline rate of the optically thin emission gets faster as shown in Appendix\,\ref{app:dependence}. Then it is needed to compensate for this faster decline rate by varying the other parameters; the choice of smaller $s$ could compensate for the fast decline introduced by smaller $n$.
This indicates the presence of a degeneracy relationship between $n$ and $s$.
That said, it remains still uncertain why our parameter estimation tends to choose the smaller $n$; it would be possible that the parameter selection is dragged by our setting of the prior functions or the model description of radio SNe, although both of them follow the widely-accepted formalisms. While we leave the preference of $n\sim5$ as one of our main results, this could be largely affected by the setting of the prior function with the lower bound set to $n=5$ 
However, this is limited by the current radio SN model using the analytical formalism for the shock expansion, and further investigating this issue will require hydrodynamic simulations.

The preference for a shallower outer ejecta density slope may result from the underestimated SN shock velocity, as smaller $n$ values yield higher velocities and thus brighter radio emission (see Appendix~\ref{app:dependence}). In fact, the shock velocity from the thin-shell approximation \citep{1982ApJ...259..302C} is known to be systematically smaller than that from the exact self-similar solution \citep{1982ApJ...258..790C,2006ApJ...651..381C}, suggesting potential systematics in our model. To account for this, we test a higher value of $f_{\rm sh}$ (Equation~\ref{eq:ue} and \ref{eq:uB}), which varies between $3/4$ and $4$ in the literature \citep{2022ApJ...938...84D}. We repeated the parameter estimation using $f_{\rm sh} = 4$ and found two main effects. First, the preferred $n$ value for SN II increases but remains lower than the prediction by \citet{1999ApJ...510..379M}; for SESNe, $n < 7$ is still favored. This supports the idea that increasing $f_{\rm sh}$ affects $n$ via shock velocity but does not qualitatively change our conclusions. Second, lower $\logee$ is favored, likely because larger $f_{\rm sh}$ compensates for $\epsilon_e$ in light curve fitting. The tendencies of other parameters remain unchanged. Hence, we conclude that the possible systematics from the thin-shell approximation do not significantly impact our result.

Indeed, the shallow density slope of the SN ejecta may be actually realized, and it could provide an important insight into the nature of the SN-CSM interaction. The widely-adopted somewhat steep slope is based on the `classical' theoretical expectation of the shock breakout from the stellar surface \citep{1999ApJ...510..379M} which does not take into account the accumulating evidence of a very dense environment in the vicinity of an exploding star (e.g., confined CSM). Then, the outer ejecta density structure can be substantially modified by the interaction with the confined CSM and subsequent `wind-breakout'. For instance, \citet[][]{2023MNRAS.521.1897M} showed that the outer ejecta structure becomes flatter if there is a massive envelope (or confined CSM) for the case of SNe Ia. This should also be realized in the case of core-collapse SNe with the confined CSM, as demonstrated by \citet{2025arXiv250414255M}, who further addressed the origin of this flat distribution in detail. Our result of the preference of small $n$ is indeed consistent with this picture, and worth further exploration in future work.

\section{Summary}\label{sec:summary}
In this study, we conduct a systematic investigation of 32 radio SNe with an MCMC analysis based on Bayesian statistics and carry out parameter fitting for each SNe. The parameters include the CSM density scale ($\logAast$), efficiencies of electron acceleration ($\logee$) and magnetic field amplification ($\logeb$), spectral index of electrons ($p$), and the density gradient of CSM ($s$) and ejecta ($n$). By utilizing the results obtained from our parameter estimation, we construct the one-dimensional marginalized probability density functions of each parameter in successfully fitted 27 radio SN samples. Our findings and insights are summarized as follows.

\begin{itemize}
    \item The distribution of the CSM density scale ($\tilde{A}_\ast$) in SNe II spreads for orders of magnitudes due to the limited number of SN II samples, while that of SESNe is clustering at $\logAast\sim2$. The mass-loss rates of SN II and SESN progenitors have been estimated for $10^{-7}\lesssim\dot{M}<10^{-4}\,M_\odot$ and $\dot{M}\sim10^{-3}\,M_\odot$, respectively, on the assumption of typical wind velocities for each SN type.
    \item The medians of $\epsilon_e$ and $\epsilon_B$ are found to be $-3\lesssim \logee \lesssim -2$ and $\logeb\simeq-2$ with large $1\sigma$ credible interval associated. The ratio of $\epsilon_e$ to $\epsilon_B$ would be diversely taken by individual radio SN samples, while the equipartition between $\epsilon_e$ and $\epsilon_B$ could not be overall ruled out.
    \item When we include early-phase data ($30\,{\rm days}$) tracing innermost CSM into our analysis, we confront the consequence that many radio SN samples tend to prefer high value of $\logeb \gtrsim -1$, which is beyond the implication of recent particle-in-cell simulations \citep{2011ApJ...726...75S, 2014ApJ...794...46C}. We take a stance that there is a hidden physical setup that causes the preference of high $\epsilon_B$ and identify it as high CSM density in the early phase. This interpretation proposes the possibility of the universality of the density enhancement of the CSM immediately near the progenitor, while the possibility of the truly high $\epsilon_B$ cannot be rejected.
    \item The index of relativistic electrons' energy spectral distribution $p$ is suggested to be greater than 2.5 for a large sample of radio SNe, indicating a softer spectrum than the prediction of pure diffusive shock acceleration theory. This suggests the possible existence of processes that softens the spectra of electrons accelerated in SNe such as nonlinear effects \citep{2010APh....33..307C, 2010MNRAS.407.1773C,2012ApJ...750..156L,2013ApJ...762L..24M}.
    \item The density slope of CSM prefers $s\simeq2$ within $1\sigma$ credible interval, corresponding to the steady wind structure. That said, many SESN samples are found to exhibit either medians or modes smaller than $2$, although the mismatch between the median and mode makes the result less robust. Since $s<2$ indicates a decreasing mass-loss rate of progenitors towards the core collapse, this may serve as a hint for understanding the mass-loss mechanism of hydrogen-deficient SN progenitors. 
    \item The density gradient of the outer layer of SN II ejecta is steeper (larger $n$) than that of SESNe. This is qualitatively consistent with the prediction in \citet{1999ApJ...510..379M}, but the values of $n$ in some radio SN samples are as small as $n\simeq5$, the lower limit of the model application. We suggest the wind-breakout from the confined CSM as a possible origin of the small value of $n$, but this result should be carefully interpreted taking possible biases and systematics into account.
\end{itemize}

Finally, it should be again noted that our radio SN samples are limited to those who have clear detections of a peak in the light curve \citep[for the other samples see][]{2021ApJ...908...75B}. Thus the results obtained in our analysis would be biased towards the SNe producing bright radio emission that is observable for us. Nevertheless, our results would provide a range of hints for understanding the nature of SN ejecta, CSM, and shock acceleration physics in core-collapse SNe.

\section*{Acknowledgements}
The authors appreciate Kai Ikuta, Kaori Obayashi, Yuri Sato, Ryo Yamazaki, Tatsuya Matsumoto, Kohta Murase, and Daichi Hiramatsu for their comments that have deepened the content of this paper.
T.M. acknowledges the support from Japan Society for the Promotion of Science (JSPS) KAKENHI grant No.~JP21J12145, the National Science and Technology Council, Taiwan under grant No.~MOST 110-2112-M-001-068-MY3 and the Academia Sinica, Taiwan under a career development award under grant No.~AS-CDA-111-M04.
K.M. acknowledges support from JSPS KAKENHI grant Nos.~JP24H01810 and JP24KK0070.
S.S.K. acknowledges support from JSPS KAKENHI grant Nos.~JP22K14028, JP21H04487, and JP23H04899, and Tohoku Initiative for Fostering Global Researchers for Interdisciplinary Sciences (TI-FRIS) of MEXT's Strategic Professional Development Program for Young Researchers.
M.T. acknowledges support from JSPS KAKENHI grant Nos.~23H04894 and 23H04891.
Parameter estimations in this study were carried out in the supercomputer cluster Yukawa-21 equipped by the Yukawa Institute for Theoretical Physics.

\software{
{\tt emcee} \citep{2013PASP..125..306F},
{\tt corner.py} \citep{2016JOSS....1...24F}
}




\bibliographystyle{aasjournal}
\bibliography{manuscript} 



\appendix

\section{A list of radio supernovae with special remarks}\label{app:anomaly}

In this appendix, we summarize radio SNe that exhibit behaviors deviating from standard model predictions. We present case studies of individual radio SNe for which specific data selection or processing approaches were necessary to analyze within our model. For each object, we explain the observational context and data considerations.

\begin{itemize}
\item SN~1987A (II): \citet{1987Natur.327...38T} first reported the radio observation of SN\,1987A at the frequency range lower than 10\,GHz. Indeed the monitoring observation had been conducted \citep{1995ApJ...453..864B,2001ApJ...549..599B} until the emergence of the outer ring \citep[e.g.,][]{2010ApJ...710.1515Z}. The property of this outer ring and the relevant environment has been investigated \citep{2015ApJ...810..168O,2019A&A...622A..73O,2020A&A...636A..22O}. However, our model is not capable of modeling this outer ring because of an assumption of a single-component CSM. In this study, we use the observational data obtained within 1\,year since the explosion, tracing the CSM formed after the outer ring formation.
\item SN~1993J (IIb): \citet{1998ApJ...509..861F} showed that SN~1993J would have dense CSM so that FFA can be important in the initial phase after the explosion. As described in Section\,\ref{sec:fitting} the initial observational data is excluded in our parameter estimation procedure and we expect that the contribution of the FFA is automatically excluded in the main result. In addition, \citet{2007ApJ...671.1959W} reported the steep decline of radio emission from SN~1993J, arguing the steep CSM density profile. Considering the large number of detection data of this object, we only treat the observational data obtained within $600\,{\rm days}$ since the explosion. 
\item SN~2001gd (IIb): \citet{2007ApJ...671..689S} reported the rapid declining of the radio luminosity after $t\gtrsim\,550\,{\rm days}$. They suggest that this temporal change is caused by the initiation of the interaction with the more tenuous CSM, similar to the case of SN\,1993J. Modeling of this rapid decline of radio luminosity is beyond the capability of our radio SN model. We exclude the observational data obtained after $t\gtrsim1000\,{\rm day}$ in order to omit the contribution from the rapid decay.
\item SN~2001ig (IIb): \citet[][]{2004MNRAS.349.1093R} reported the radio observation of SN\,2001ig, which exhibits the modulations in the light curve. They suggest the scenario that the modulation is produced by the interaction between the CSM, which is bent by the orbital motion of the Wolf-Rayet star in the binary. We use the observational data obtained within the time range of $t\lesssim 70\,{\rm days}$, tracing only the first peak in the light curve to make the statistics more robust. In this case, the proposed parameter set is tracing the CSM in the vicinity of the progenitor, or immediately before the explosion. We note that, we attempted the parameter estimation using all of the obtained data, but the resultant chi-square score was not good ($\chi_{\rm red}^2=5.9$). In this case, we found that $\epsilon_B\sim1$ has been proposed as a best-fitted parameter, but this is implausible from the viewpoint of the magnetic field amplification efficiency.
\item SN~2003bg (IcBL): \citet[][]{2006ApJ...651.1005S} reported the radio observation of SN~2003bg, exhibiting the modulation of the radio luminosity. Their modeling suggests the enhancement of the CSM density that can be interpreted as the temporal evolution of the mass-loss rate of the progenitor. We use the observational data obtained within the time range of $t\lesssim 100\,{\rm days}$, tracing only the first peak in the light curve to make the statistics more robust. Similar to the case of SN~2001ig, the parameter estimation with all of the observational data included results in the bad chi-square score ($\chi_{\rm red}^2 = 7.03$).
\item SN~2004cc: \citet{2012ApJ...752...17W} reported the radio observation of SN~2004cc, and they claimed that this SN exhibited the re-brightening of the radio emission at $t\sim135\,{\rm days}$ after the explosion. We excluded this re-brightening data in our parameter estimation. Note that a similar exclusion has been done in \citet{2012ApJ...752...17W}.
\item SN~2004dk (Ib): \citet{2012ApJ...752...17W} reported the radio observation of SN~2004dk, and they claimed that this SN also exhibited the re-brightening of the radio emission at $t\sim4\,{\rm years}$ after the explosion. We excluded this re-brightening data in the parameter estimation procedure. Note that a similar exclusion has been done in \citet{2012ApJ...752...17W}.
\item SN~2007bg (IcBL): \citet[][]{2013MNRAS.428.1207S} reported the radio observation of SN~2007bg, exhibiting the drastic radio rebrightening $t\sim1\,{\rm year}$ after the explosion. They proposed the scenario that the CSM density enhancement would explain the late-time re-brightening of radio emission. In our parameter estimation, we use the observational data obtained within $t\lesssim 200\,{\rm days}$ after the explosion, tracing only the first peak of the light curve.
\item SN~2007uy (Ib): \citet[][]{2011ApJ...726...99V} reported the radio observation of SN~2007uy with a range of the observational frequency from 325\,MHz to 8.4\,GHz. They found the excess of radio emission at 325~MHz and attributed it to the appearance of the off-axis jet. Given this, we do not include the observational data at 325~MHz in our parameter estimation.
\item SN~2014C (Ib): \citet[][]{2017MNRAS.466.3648A} reported the radio monitoring of SN\,2014C at the frequency of 15.7\,GHz and \citet[][]{2017ApJ...835..140M} and \citet[][]{2018MNRAS.475.1756B} reported the detection of 8.4 and 22.1\,GHz radio emission in the late phase. Combined with the results of multi-wavelength observations, it has been considered that SN~2014C possesses a dense CSM detached from the progenitor, producing bright observational signatures in a wide range of wavelengths. In this study we use the radio observational data obtained within $t\lesssim200\,{\rm days}$ after the explosion, only tracing the first peak of the radio light curve, similar to the case of SN~2007bg.
\end{itemize}

Next, we mention some SN objects that we do not incorporate into our MCMC analysis.

\begin{itemize}
\item SN~1998bw (IcBL): As reported in \citet{1998Natur.395..663K}, the radio light curve in SN~1998bw is characterized by double peaks. A double-peaked feature cannot be reproduced in the standard model for radio SNe based on one component of the CSM. We find that the first peak exhibits within 30\,days since the explosion, which would not be treated in our data processing. Thus we do not attempt the fitting of SN\,1998bw.
\item SN~2004C (IIb): \citet[][]{2022ApJ...938...84D} reported the long-term radio observation of SN~2004C, exhibiting the bright radio emission a year after the explosion. They claim that this radio emission is characterized by a shell-like CSM, rather than a steady wind. Indeed our parameter estimation has failed to find out the feasible parameter set that can reproduce the radio light curve of SN\,2004C. Therefore we do not include this object in our analysis.
\item SN~2008ax (IIb): \citet[][]{2009ApJ...704L.118R} reported the radio observation of SN~2008ax, which showed the enhancement of the radio luminosity at the age of $t\sim60\,{\rm days}$. Similar to the case of SN~1998bw, the observational data containing the first peak consists of only the data obtained within 30\,days since the explosion, which would be removed by our criteria. Thus we do not attempt the fitting of SM\,2008ax.
\item PTF11qcj (IcBL): \citet[][]{2014ApJ...782...42C} reported the radio observation of a transient PTF11pcj, and its temporal evolution of radio emission is complicated. Our parameter estimation has failed to find the feasible parameter set that can reproduce the light curve of PTF11qcj, and we do not include this object in our analysis.
\end{itemize}

\section{Fitting results of each SN with the early-phase observational data included}\label{app:allepoch}

\begin{figure*}[htbp]
\includegraphics[width=0.5\linewidth]{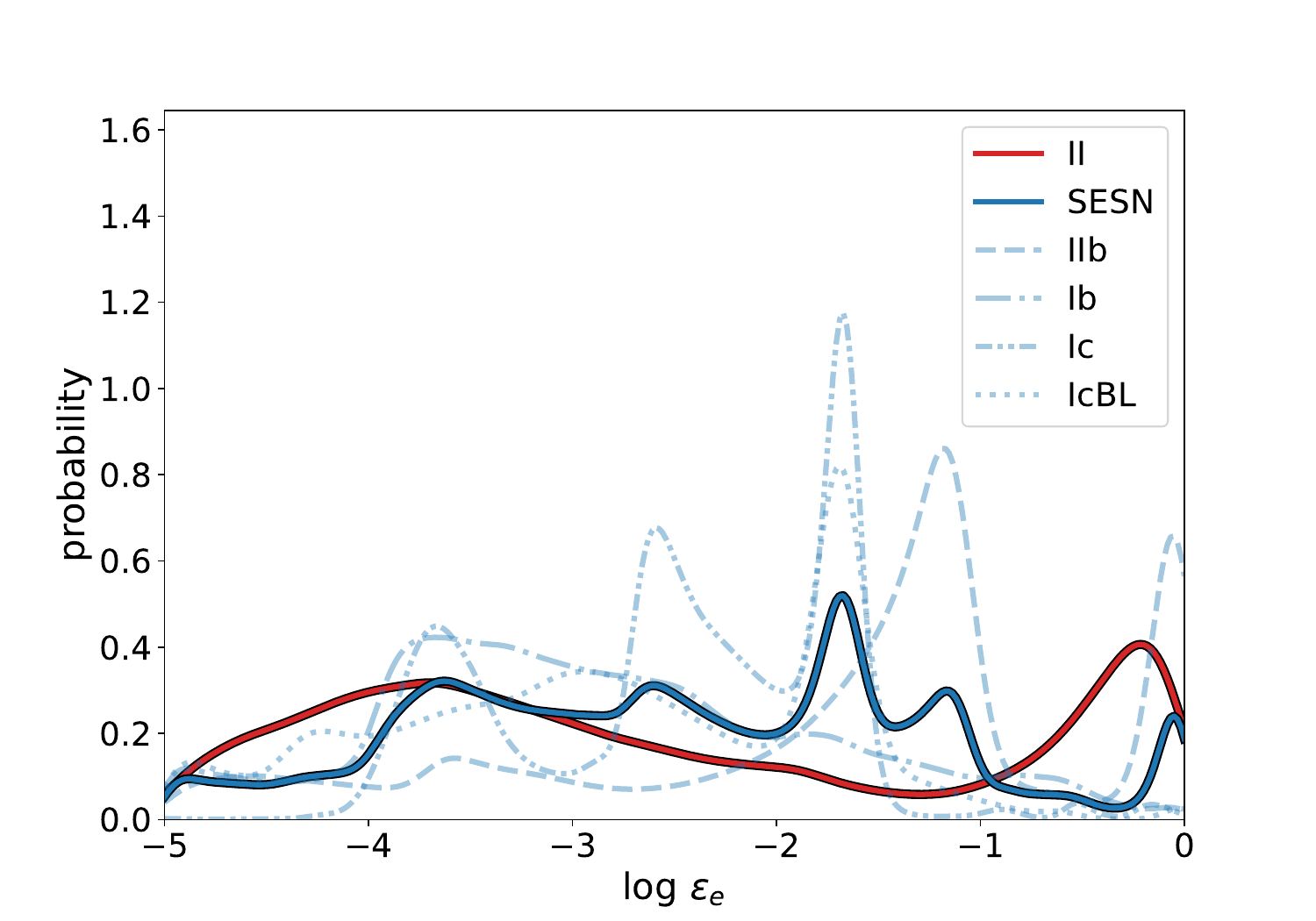}
\includegraphics[width=0.5\linewidth]{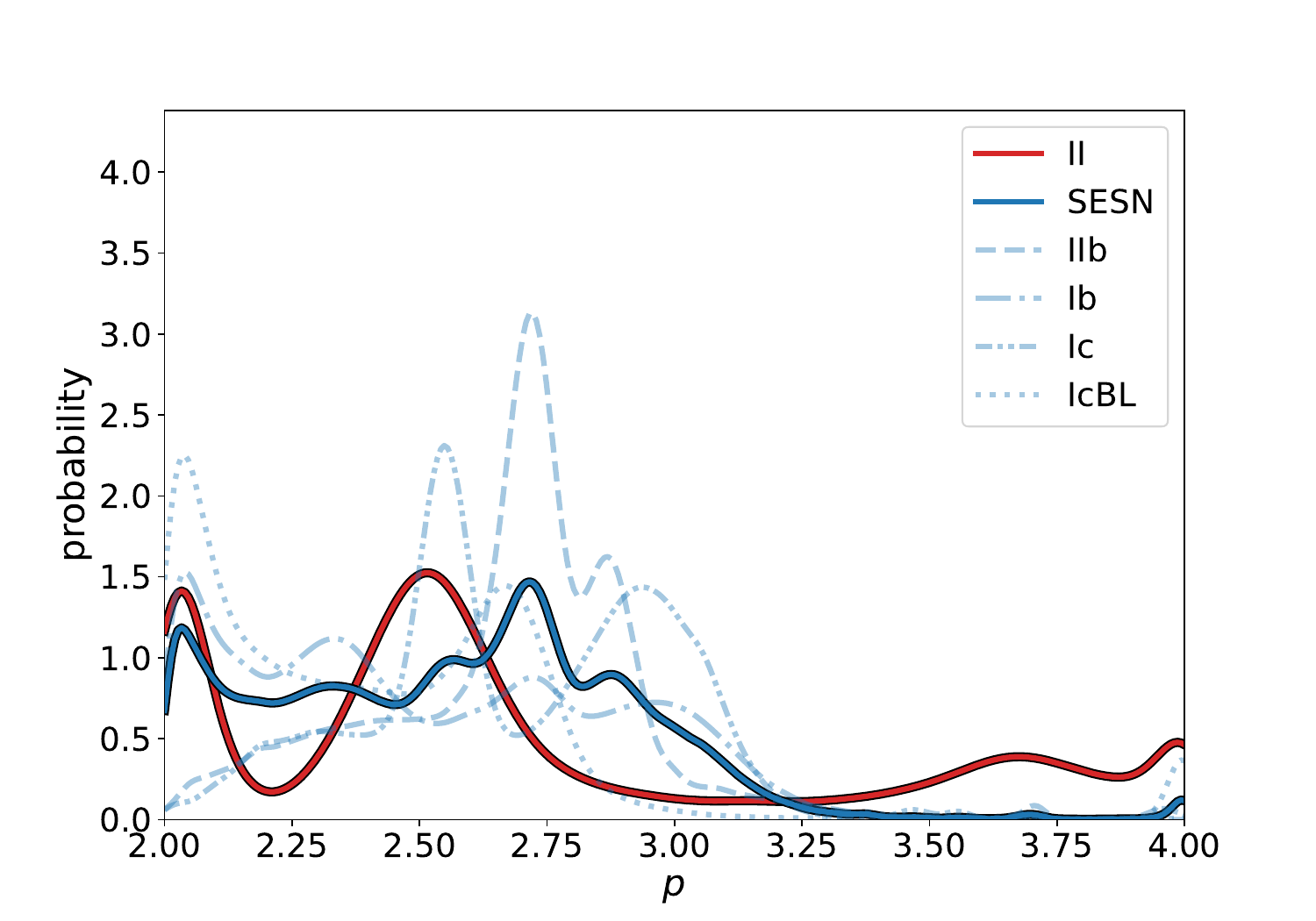} \\
\includegraphics[width=0.5\linewidth]{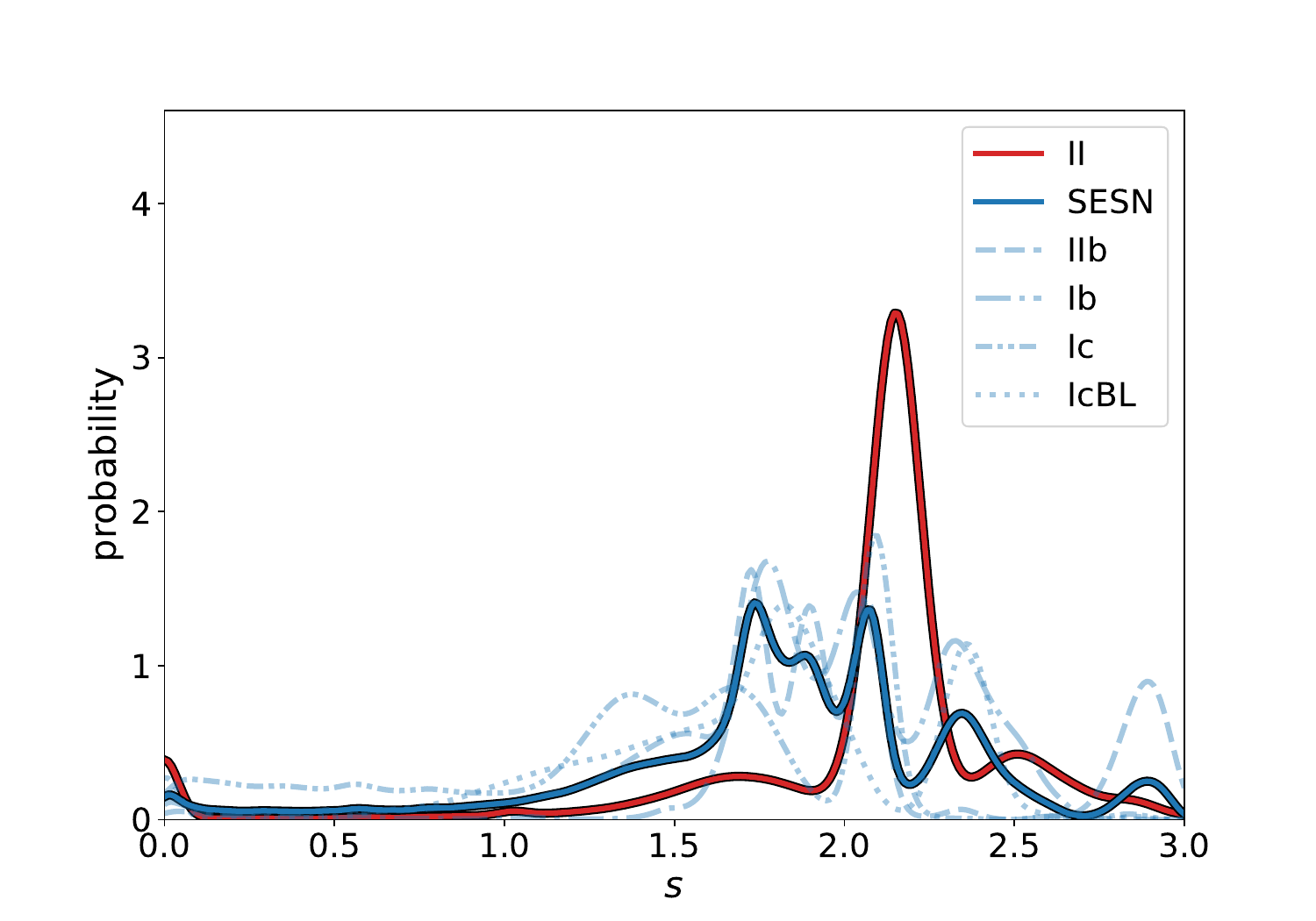}
\includegraphics[width=0.5\linewidth]{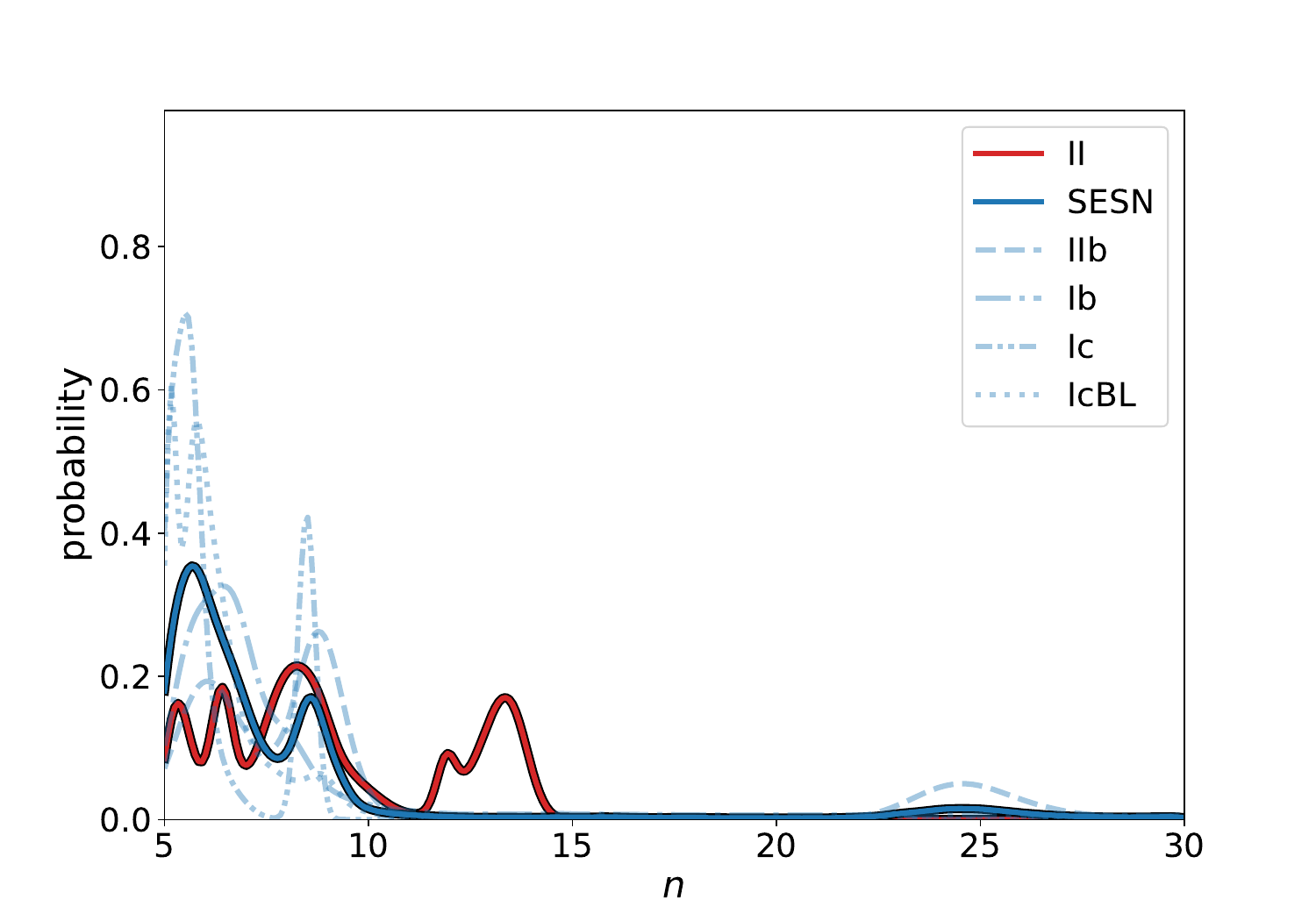}
\caption{The marginalized one-dimensional probability density functions of (a) $\logee$, (b) $p$, (c) $s$, and (d) $n$ constructed from the MCMC sample data with the early-phase observational data included.}\label{fig:PDf55}
\end{figure*}

\begin{figure}
    \includegraphics[width=0.5\linewidth]{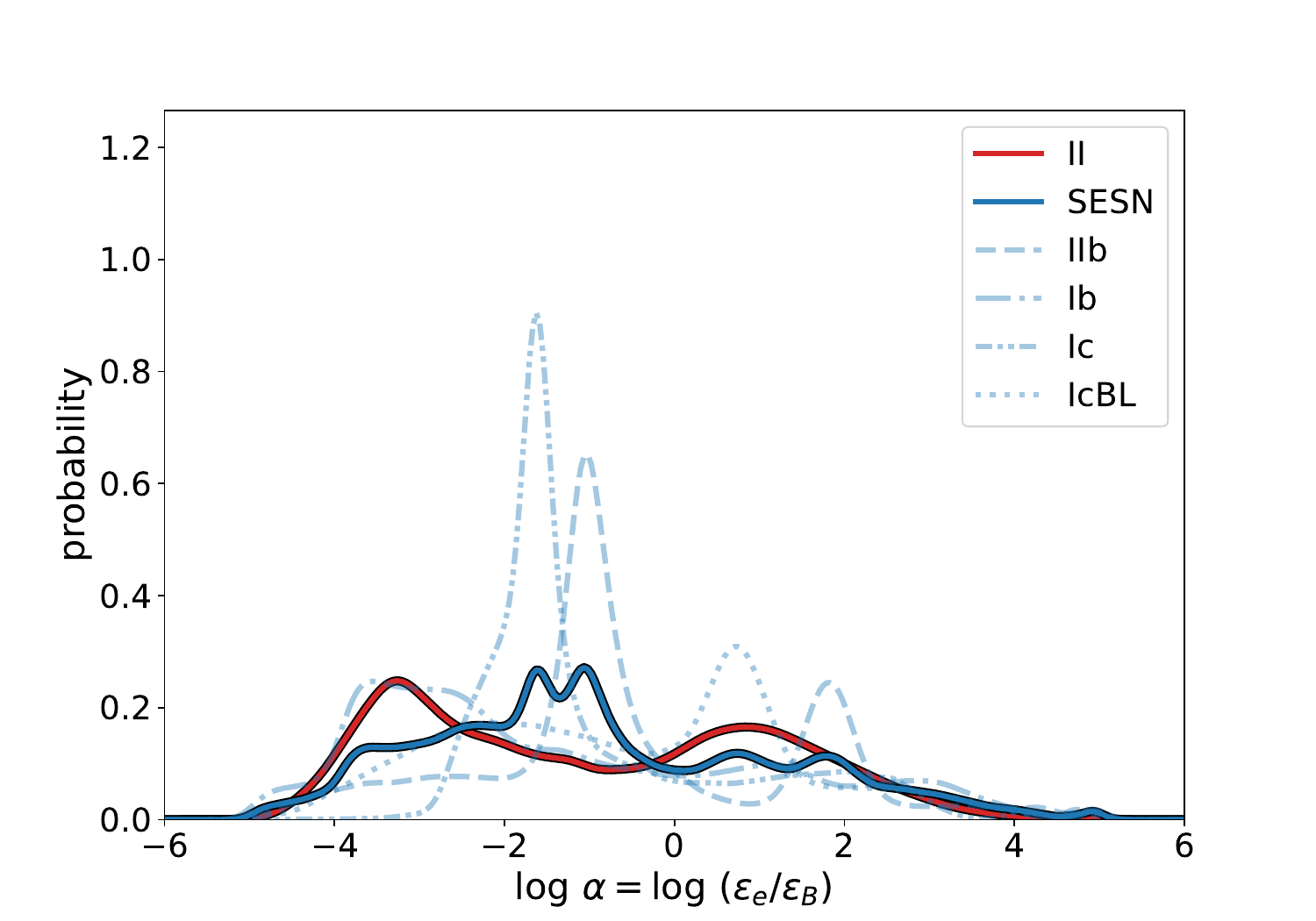}
    \includegraphics[width=0.5\linewidth]{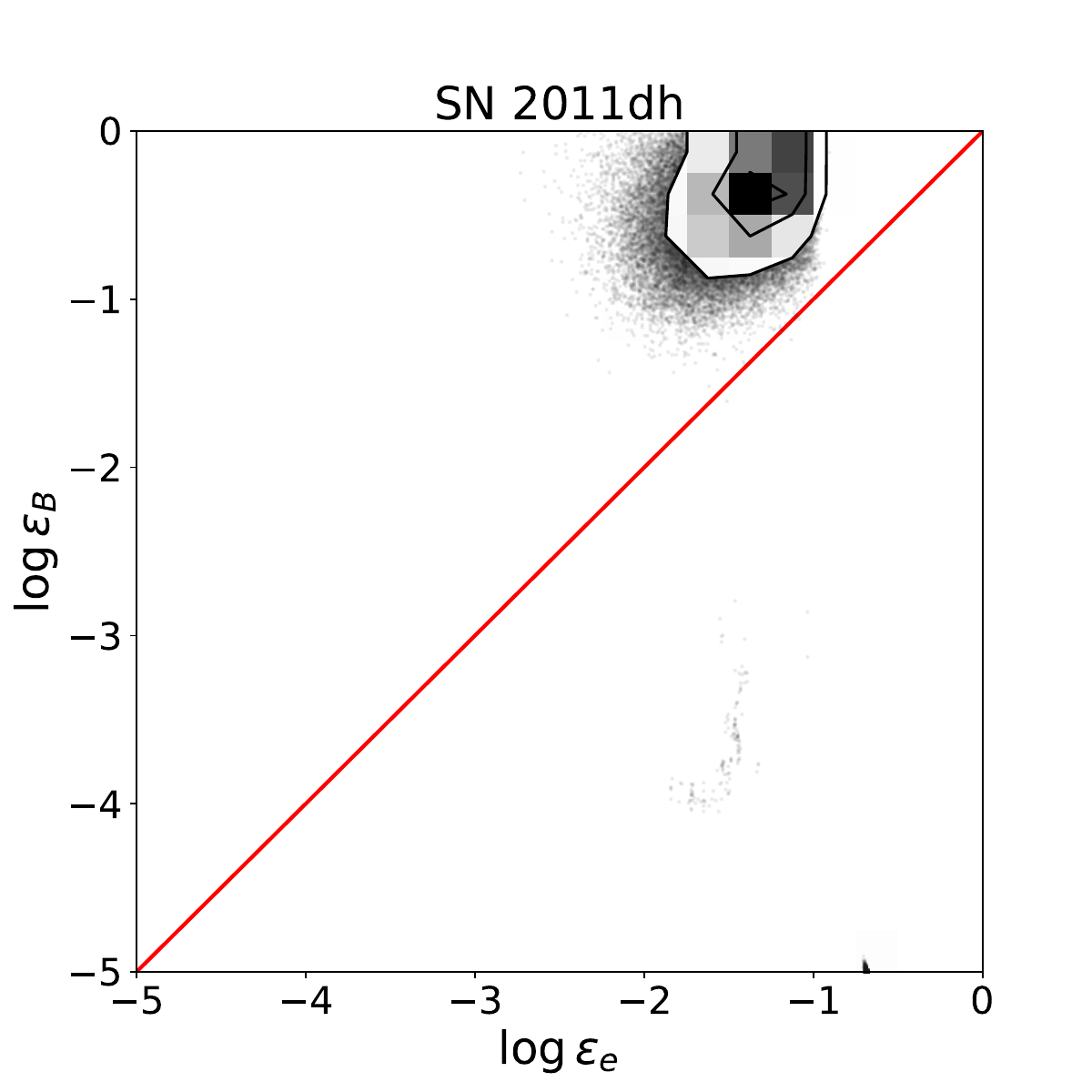}
    \caption{Same as Figure\,\ref{fig:epsilon}, but constructed from the MCMC sample data with the early-phase observational data included.}
    \label{fig:app_epsilon}
\end{figure}

In the main analysis, we used observational data without early-phase data, but it would be useful to briefly show the result of parameter estimation with the early-phase data included.
Figure\,\ref{fig:PDF_logA_logeB_early} showed the probability density functions of $\logAast$ and $\logeb$, and those of the other parameters are shown in Figure\,\ref{fig:PDf55}.
In addition, it would be useful for readers to visualize whether the equipartition between relativistic electrons and magnetic field is satisfied even if the early-phase data is included.
This is shown in the left panel of Figure~\ref{fig:app_epsilon}, in which we can see the preference of $\log\alpha$ less than 0, indicating the predominance of the magnetic field over electrons.
This result is coming from the preference of extremely high $\epsilon_B$ that has been seen in many radio SN samples, whose example is shown in the right panel.
Given this, we need to consider the possibility of the systematics hidden in our framework as discussed in Section~\ref{subsec:CSM_vicinity}.

\section{Dependence of a radio light curve on the surveyed parameters}\label{app:dependence}

\begin{figure}
    \includegraphics[width=\linewidth]{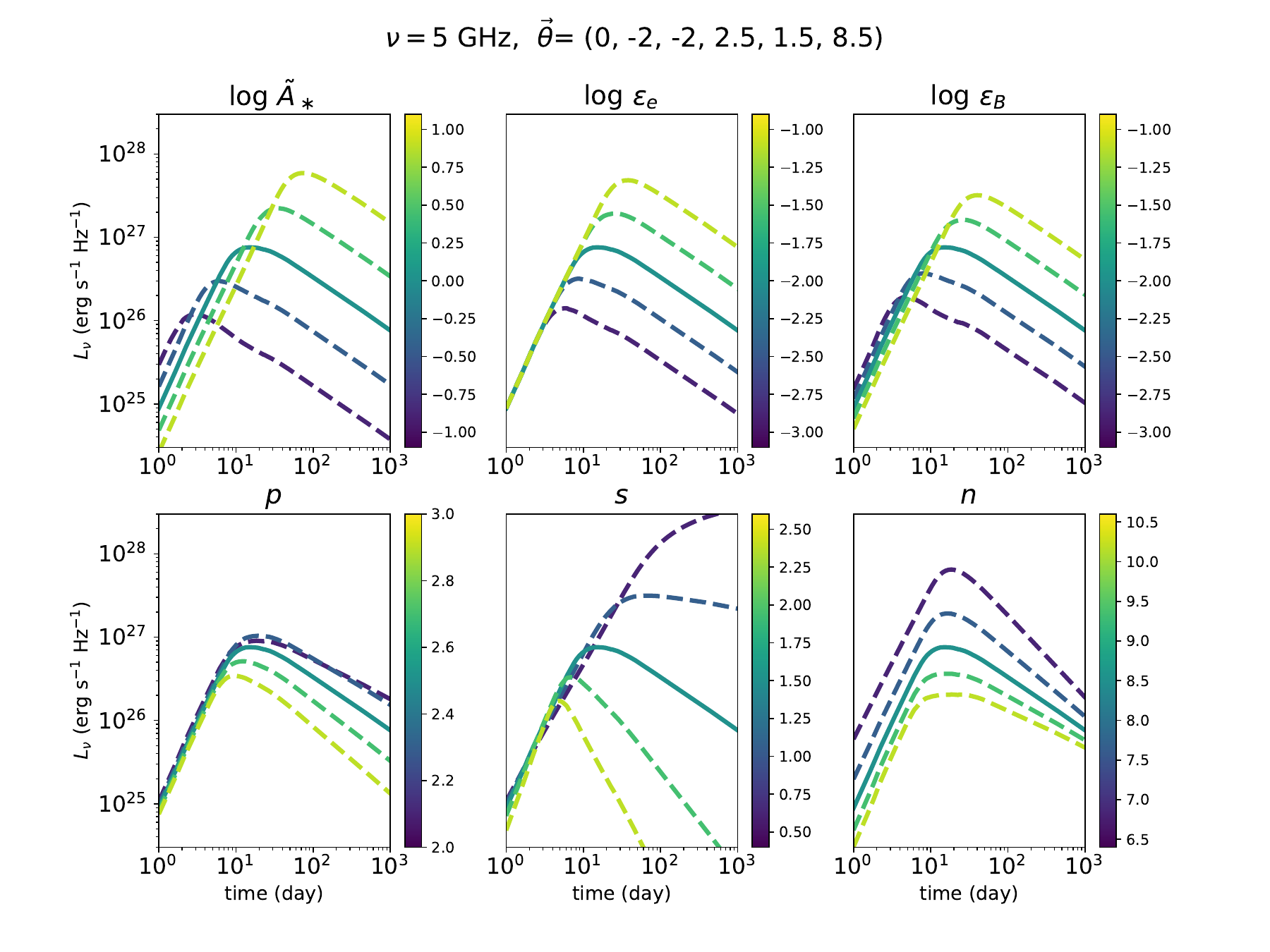}
\caption{Parameter dependences of 5~GHz radio light curve reproduced by our radio SN model. The solid lines in all panels are the benchmark model with $\logAast=0, \logee=-2, \logeb=-2,p=2.5,s=1.5$, and $n=8.5$. In each panel, only the parameter on the title is varied according to the accompanied color bar from the reference model, whose light curves are plotted by the dashed lines.}
    \label{fig:app_dependence}
\end{figure}

In this section, we show the parameter dependence of the radio light curve reproduced from our model, which is shown in Figure\,\ref{fig:app_dependence}. As a benchmark model, we employ the ejecta property and bolometric light curve of SN\,2002ap and we use the parameter set of $\logAast=1,\logee=-2,\logeb=-2,p=2.5,s=1.5,n=8.5$. This parameter set is not tuned to the actual model of SN\,2002ap but is just used for convenience. The benchmark model at 5\,GHz is plotted by a solid line. The other models are plotted by the dashed line, in which the parameter focused in each panel is varied in the range of the color bar next to the panel while the other parameters are still fixed. Note that the bump seen around $t\sim30\,{\rm days}$ is attributed to the inverse Compton cooling with the seed photon originating from the bolometric light curve \citep{2013ApJ...762...14M}.

First, we elaborate on the dependence on $\tilde{A}_\ast, \epsilon_e$, and $\epsilon_B$, all of which mainly influence the bulk value of the peak time and peak luminosity. The noteworthy point is that the increase in $\tilde{A}_\ast$ and $\epsilon_B$ results in the same behavior characterized by the dimming optically thick emission and the brightening optically thin emission. While the increase in $\logeb$ reacts to only the strengthening of the magnetic field, the increase in $\logAast$ in addition leads to the increase in the population of relativistic electrons. As a result, these two parameters enlarge the optical depth to synchrotron self-absorption, resulting in the delay of the peak time and the brightening of the peak luminosity.

The dependence of $\logee$ looks similar to those of $\logAast$ and $\logeb$, but the critical difference is that the increase in $\logee$ does not change the optically thick emission. $\logee$ adjusts the population of relativistic electrons. The important property is that both the emissivity and self-absorption coefficient linearly depend on the normalization of the electron SED; when we calculate the source function of synchrotron emission, the dependency on $\epsilon_e$ will be canceled out. Hence, the change of $\logee$ only appears in the luminosity in the optically thin phase. This property is taken into account when we attempt to infer the CSM environment in the vicinity of the progenitor (see Section\,\ref{subsec:CSM_vicinity}).

Next, we focus on $p$ that controls the spectral index of the energy distribution of relativistic electrons. This parameter mainly determines the spectral slope of the observed synchrotron emission and less affects the light curve shape. The variation we observe is that as $p$ gets smaller (the SED becomes harder) the optically thin emission gets brightened. The interpretation would be straightforward; with the hard electron SED given the number of electrons with high Lorentz factor increases, which largely contribute to the synchrotron emission at the frequency we are interested in. On the other hand, the optically thick emission involves subtle variation because the information of the electron SED is canceled out in the optically thick emission, as described in the case of $\epsilon_e$.

$s$ and $n$ serve significant variations in the light curve, unlike the case of $p$, because these parameters adjust the density slopes of the CSM and ejecta, respectively. The variation of $s$ affects both the SN shock velocity and the CSM density swept by the shock. As a result, the slowly evolving light curve is reproduced in the case of smaller $s$. The variation of $n$ determines only the temporal evolution of the SN shock. Since the smaller $n$ gives rise to the rapid evolution of the SN shock, the resultant light curve exhibits a rapid decline rate with time. We also note that smaller $n$ reproduces the brighter peak luminosity because it makes the shock velocity higher.

\end{document}